\documentclass[a4paper,amsmath,amsfonts]{quantumarticle}
\pdfoutput=1
\usepackage{graphicx}
\usepackage{color}
\usepackage[colorlinks=true,citecolor=blue,linkcolor=blue,urlcolor=blue]{hyperref}

\usepackage{braket}

\def\echo{Loschmidt echo }

\def\Ts{{\mathcal T}}

\def\bz{{\bar{z}}}
\def\bh{{\bar{h}}}
\def\bw{{\bar{\omega}}}

\newcommand{\Eq}[1]{{Eq.~({\ref{#1}})}}
\newcommand{\Fig}[1]{{Fig.~{\ref{#1}}}}

\newcommand{\beq}{\begin{equation}}
\newcommand{\eeq}{\end{equation}}

\graphicspath{{./}{./figures/}}

\begin{document}

 \title{Loschmidt echo,  emerging dual unitarity  and scaling of  generalized temporal entropies  after quenches to the critical point}
\author{Stefano Carignano}
\affiliation{Barcelona Supercomputing Center, 08034 Barcelona, Spain}
\email{stefano.carignano@bsc.es}
\author{Luca Tagliacozzo}
\affiliation{Institute of Fundamental Physics IFF-CSIC, Calle Serrano 113b, Madrid 28006, Spain}
\email{luca.tagliacozzo@iff.csic.es}

\begin{abstract}
{We show how} the \echo of a product state after a quench to a conformal invariant  critical point and its leading finite time corrections can be predicted by using conformal field theories (CFT). We check such predictions with tensor networks, finding excellent agreement.
As a result, we can use the \echo to extract the universal information of the underlying CFT including the central charge, the operator content, and its generalized temporal entropies.  We are also able to predict and confirm an emerging dual-unitarity of the evolution at late times, since the spatial transfer matrix operator that evolves the system in space becomes unitary in such limit.
Our results on the growth of temporal entropies also imply that, using state-of-the art tensor networks algorithms, such calculations only require resources that increase polynomially with the duration of the quench, thus providing an example of numerically efficiently solvable out-of-equilibrium scenario.
\end{abstract}
\maketitle

\section{Introduction.}
Strongly-correlated quantum systems out-of-equilibrium still challenge our understanding due to the exponential complexity involved in making predictions for them. This complexity has far-reaching implications, such as the ongoing debate surrounding the existence of the many-body localized phase, a problem that demands simulations of large systems over long time scales \cite{jccmp2023,chandran2024}.

As a result, currently most of our understanding of the out-of-equilibrium dynamics emerges either from short-time numerical simulations or from scenarios where analytical calculations can be performed. This is possible e.g. in integrable models \cite{caux2013,essler2023,bertini2016,castro-alvaredo2016}, conformal field theories \cite{calabrese_2005,calabrese_2007,stephan2011a,cardy_2016,dubail2017,surace_2020} and Floquet systems such as random  \cite{nahum2017} or dual unitary circuits \cite{bertini2018,bertini2019,bertini2019a,claeys2020,claeys2021}.

The relation between these classes is still not completely clear.
For example, dual unitary circuits are quantum circuits that describe the Floquet dynamics of a system driven by local interactions encoded in elementary gates, which possess an underlying spacetime duality that results in a circuit encoding a unitary dynamics in both time and space. 
Dual unitary gates have also been used to construct toy models for holography which display emerging discrete Lorentz and conformal invariance \cite{masanes2023}, hinting to a possible connection between dual unitarity and conformal invariance.


In this work, we start from Hamiltonian dynamics and focus on the Loschmidt echo, namely 
 the return probability of an initial product state, after a quench to the critical conformal invariant point.
The behavior of such return probability provides key information of the dynamics and can be used as a probe of quantum chaos since its behavior can be related to that of out-of-time ordered correlations \cite{yan2020} (for earlier studies see also \cite{pozsgay2013,andraschko2014,piroli2017}, for a review see \cite{goussev2012}).  
We map the problem into a path integral calculation of a conformal field theory (CFT) on a strip where the boundary correspond to the initial state. Such boundary CFT (bCFT) can be solved using the powerful machinery of conformal maps, which allows us to obtain predictions about the leading \echo exponential decay at large times. We unveil that the leading decay is dictated by the central charge of the CFT \cite{cardy_1984,affleck1986}, which measures the number of degrees of freedom of the theory \cite{cardy2010a}. We also  show how the finite-time corrections to this leading exponential decay depend on the initial state, and are dictated by the operator content of the corresponding bCFT \cite{cardy1986,cardy1986a}.
We find that 
 the late-time \echo of a system described by CFT gives rise to a unitary transfer matrix in space, providing an example of emerging dual-unitarity at late times.

Finally, we define and characterize generalized temporal entropies, that arise by studying reduced transition matrices corresponding to a time-like cut in the path-integral.
 These complex-valued entropies are known to carry geometric information in holographic field theories where they have first been proposed \cite{narayan2015,narayan2016,nakata2021,doi2023}
 but their interpretation in a quantum information context is still unclear. 
   Our results for such generalized temporal entropies in Eq. \eqref{eq:Sgen_cft} are consistent with the predictions obtained from holography and alternative CFT approaches   \cite{narayan2015,narayan2016,nakata2021,doi2023,doi2023a,narayan2023,li2023},
 and show that for our setup they only grow logarithmically in time, as opposed to the standard entanglement entropies which grow linearly in time. 
 As a result, we can use recently devised tensor network (TN) algorithm based on temporal matrix product states in order to cross-check such prediction for very large times and systems in the thermodynamic limit.

We perform such TN simulations using  our recently introduced algorithm\footnote{For similar algorithms see also \cite{banuls2009,muller-hermes2012,hastings2015,lerose2021,lerose2023}.} \cite{carignano2024b}  on two exemplary minimal models, the Ising and the three-states Potts model, finding perfect agreement with the CFT predictions.

\section{Setup and CFT predictions.}

\begin{figure}
\begin{center}
 \includegraphics[width=.9\columnwidth]{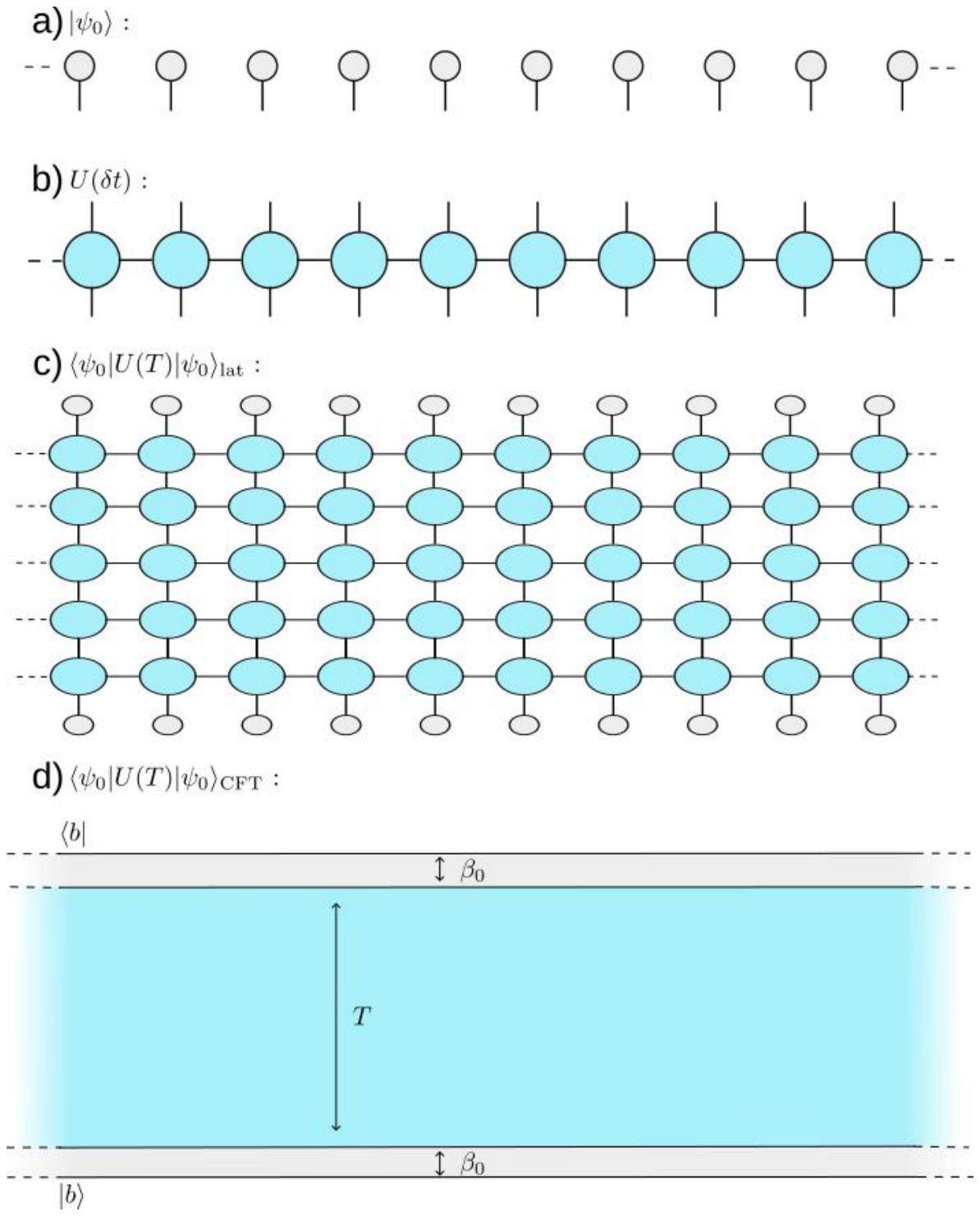}
 \caption{The \echo can be studied numerically by defining the tensor network made by a) the initial product state b) the MPO of the evolution operator and c) its matrix element. It can also be characterized using field theories, by analytically continuing the results obtained by mapping the CFT on the plane to the finite geometries shown in d).
 The field theory infrared divergence is handled by studying a finite spatial extent $L$ that is then sent to infinity, while the UV divergencies are cured by introducing a finite $\beta_0$ close to the conformally invariant boundary states $\ket{b}$.
 \label{fig:CFT} 
 }
 \end{center}
\end{figure}

Given a 1D quantum many-body system made by $L$ constituents and described by a Hamiltonian $H$, we focus on the contraction of a TN encoding the \echo (namely the return probability of a state of a system to its initial state after the out-of-equilibrium evolution for a time $T$) and its intensive part,
\begin{equation}
\mathcal{L}= |\bra{\psi_0}e^{-iHT}\ket{\psi_0}|,\quad l=- \frac{\log{\mathcal{L}}}{TL}. \label{eq:Lochmidt}
\end{equation}

Our setup is described in Fig.~\ref{fig:CFT}(a-c):
through simple tensor decompositions, we can express the evolution for an infinitesimal time step $U(\delta t)$  as a matrix product operator (MPO) with finite bond dimension (see e.g. \cite{mcculloch2007,pirvu2010}). 
As a result, Eq.~\eqref{eq:Lochmidt} 
 is given by the contraction of a regular 2D TN made by
 $L\times N_T$  $(N_T = {T}/{\delta t})$ elementary four-leg tensors, one for every space-time point. 
Since everything is translationally invariant, we can directly consider the thermodynamic limit by sending $L\to \infty$, thus building an infinite strip with a width of $N_T$ tensors in the time direction.

We now focus on critical dynamics and use the machinery of CFTs.
This is done by mapping our time evolution problem to an equivalent path integral formulation, using well-established prescriptions that allow to connect with statistical physics and field theory (see Appendix). In analogy to those systems, the relevant object here will be the transfer matrix~\cite{wang1997,sirker2005a,sirker2012}, which generates spatial translations.
 In the CFT, the initial states are conformally invariant boundary states $\ket{b}$ \cite{cardy1986a,stephan2011a} and the \echo is described by the geometry illustrated in Fig.~\ref{fig:CFT}(d), representing an infinitely long strip with boundary conditions given by the chosen $\ket{b}$. On the lattice, we accordingly  evolve an initial product state corresponding to  $\ket{b}$ for some short amount of euclidean time  $\beta_0$ \cite{calabrese_2005,calabrese_2007,stephan2011a,robertson2022}, leading to a transverse size  {$T+2\beta_0$} for our strip.
  The \echo for a  product state is obtained by taking the limit $\beta_0\to 0$.

Following well-established prescriptions \cite{calabrese_2005,calabrese_2007,robertson2022},  we obtain predictions for the real-time calculations by performing an analytic continuation of the results of CFTs in Euclidean space. 
We start with the standard results  on a strip 
with extent $\beta$ and analytically continue them to the case  $\beta \to i T+2\beta_0$ \cite{calabrese_2005,doi2023,doi2023a,guo2024}.
The  CFT predictions at finite $\beta$ are obtained by using conformal maps\footnote{Explicit CFT predictions for the \echo after local quenches have been obtained in \cite{stephan2011a}.}: The infinite strip described by complex coordinate $w$ related to the upper 2D plane by the map
 $w=\tfrac{v\beta}{\pi}\log(z)$, where $v$ is the sound velocity.

The transfer operator $\Ts$ produces translations along the strip, and can be expressed as an integral of the stress-energy tensor of the CFT.
For our strip geometry, after analytic continuation for real time and keeping only terms at first order in $\beta_0/T$, with $\beta_0 \ll T$, we find that 
the TM describing our quench is given by
\begin{align}
\Ts & = \exp{\left[-i\left(-\frac{\kappa}{T v}+\frac{\pi}{T v} L_0\right)\left(1 + 2i\frac{\beta_0}{T}\right)\right]}\,, \label{eq:tm_cyl_T}
\end{align}
where we  dropped the higher order terms $\mathcal{{O}}(1/(Tv)^2)$ vanishing in the large $T$ limit.
This equation is our bridge connecting the \echo with the universal quantities of the underlying CFT: here 
${\kappa}= -{\pi c \delta t}/{24}$,
$L_0$ is the relevant generator of the Virasoro algebra and $c$ is the central charge \cite{belavin1984,cardy1988,ginsparg1988,francesco1997,henkel1999,cardy2008}.

If we now define $\{\tau_i\}$ the eigenvalues of $\Ts$ (with $|\tau_0| > |\tau_1| > \dots$) and $\lambda_i=\log(-\tau_i)$,  one can see that the intensive part $l$ of the \echo is dictated by $\lambda_0$, the logarithm of the leading eigenvalue of $\Ts$, as $l$ converges to $ -\tfrac{|\lambda_0|}{T}$ exponentially in the thermodynamic limit for a gapped $\Ts$.
 \Eq{eq:tm_cyl_T} identifies the spectrum of $\Ts$ with that of $L_0$, which is diagonal in the basis of scaling fields. As a result, the CFT  predicts the full spectrum of $\Ts$, which is given by the operator content of the corresponding CFT with boundaries \cite{cardy1986, cardy1986a}.

To make contact with the lattice results, using standard procedures we shift the set of $\set{\lambda_i}$ by two
non-universal terms $a \beta + b$ ($\beta$ being again the width of the strip), that encode the normalization of the Hamiltonian and boundary effects  \cite{cardy_1984,affleck1986}. 
For the dominant eigenvalue $\lambda_0$, after analytical continuation we thus obtain at leading order 
%
%
%

  \begin{equation}
  Re(\lambda_0) = 2\beta_0 a v + b \,,\quad
  Im(\lambda_0)=  avT - \frac{\kappa}{vT} \,,
  \label{eq:lambda0exp}                                                                                                                                                                                                                              
  \end{equation}
%
whereas for the gaps of $\Ts$ we find
$Re(\lambda_1 - \lambda_0) = 0$ and 
\beq
Im(\lambda_1 - \lambda_0) =  - \frac{p\pi x_1} {vT} \,, 
\label{eq:gapsexp}
\eeq
%
%
%
where $x_1$ is the smallest boundary critical exponent. 
The leading behavior of the imaginary parts of the eigenvalues of the TM for our \echo is thus determined by the universal quantities of the underlying CFT, namely
the central charge and the critical exponents.

Our predictions extend to the whole spectrum of $\Ts$,
dictating that all its gaps
 shrink to zero as $T$ increases, 
 and are determined by the various 
 critical exponents of the model, 
 which are selected by the boundary conditions  \cite{cardy1986a,affleck1998}.
 As a result, by studying the decay of \echo as a function of time, we have access to a full set of universal properties of the underlying critical dynamics, including 
 the central charge and critical exponents.
 
These results also unveil a connection between the CFTs and dual unitary evolutions: up to higher-order corrections, which vanish as $T$ increases,  $\Ts$ has full sectors of gaps that are purely imaginary, and collapse on the unit circle for $\beta_0 \to 0$. This suggests that $\Ts$ becomes unitary in the limit of $T \to \infty$, something we refer to as an \textit{emerging dual unitarity} of the dynamics in the large-T limit.

We now focus on the computational complexity of the evolution, that is dictated by the growth of the temporal entropy. 
Such growth can be obtained from CFT: in particular, from the two-point correlation functions of twist fields 
we can access the Tsallis entropies of order $n$ \cite{srednicki1993,callan1994,vidal2003b,calabrese2004,caraglio2008}.
For a time-like separation along the imaginary axis, they correspond to the traces of the $n$-th power of the reduced transition matrices, defined as
\begin{equation}
 \mathcal{T}^{L|R}_t = \textrm{tr}_{N_t-t} \left[ \mathcal{T}^{L|R}\right]  \,,\quad
 \mathcal{T}^{L|R} =\frac{\ket{R}\bra{L}}{\braket{L|R}},
  \label{eq:rtm}
\end{equation}
where $\ket{R}$ and $\bra{L}$ are the right and left dominant eigenvectors of $\Ts$.
By continuing 
these results
to $n=1$  we obtain the results for the generalized entanglement entropies \cite{nakata2021} of  $\ket{R} $ and $\bra{L}$, which in the limit of small $\beta_0$
becomes
\begin{equation}
S_{gen} = s_0 + \frac{i \pi c}{12} + \frac{c}{6} \log \Big[ \frac{2 T}{\pi} \sin \Big( \frac{\pi t}{T} \Big) \Big] \,,
\label{eq:Sgen_cft}
\end{equation}
with $s_0$ a constant.

Notice that even though  $\Ts$ becomes unitary, the leading eigenvectors only have logarithmic generalized entropies. This fact is due to the CFT Hilbert space structure, whose number of states (at low energy) only diverges polynomially with $T$. The small finite $\beta_0$ project onto that subspace. Similar effects have been observed in \cite{robertson2022,grundner2023,cao2024}.

Drawing analogies with the CFT predictions for the spatial entropy of grounds state of critical quantum chains, we can associate the logarithmic growth of the 
generalized entropies to the closing of a gap in the transfer matrix spectrum. As we move away from the critical point, we expect this gap to open again, resulting in an area law for the generalized temporal entropies due to the finite correlation length of the system:  $S_{gen} \sim \frac{c}{6} \log(\xi)$, $\xi$ being proportional to the inverse of the gap. In turn, this property guarantees an even more efficient representation of the dominant vectors of the TM in terms of tMPS.

\begin{figure}
\begin{center}
  \includegraphics[width=.9\columnwidth]{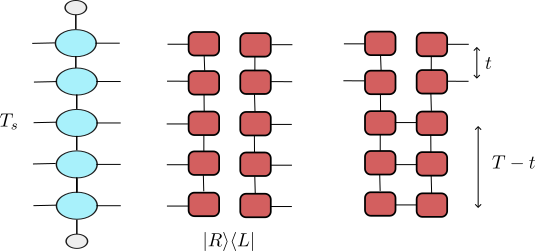}
 \caption{The transverse transfer matrix $\Ts$ involves the contraction of a column of elementary tensors a). Its leading left and right eigenvectors b) allow to characterize the \echo \eqref{eq:Lochmidt} in the thermodynamic limit. Generalized temporal entropies are extracted from  the reduced transition matrices defined in Eq. \eqref{eq:rtm}.
 \label{fig:tMPO}}
 \end{center}
\end{figure}

We can now check the above results using matrix product states (MPS). We define a transfer-matrix temporal MPO (tMPO) $\Ts$ by contracting a vertical strip of tensors, and find its leading left and right eigenvectors in the form of temporal MPS (tMPS) \cite{banuls2009,muller-hermes2012,hastings2015,tirrito2022,lerose2021,lerose2023} as shown in Fig.~\ref{fig:tMPO}. 

\section{Numerical results.}
We consider both the transverse field Ising model with Hamiltonian
\begin{equation}
{{H}_{Ising}}(g) = -  \sum_i  \Big[ \sigma_x^i \sigma_x^{i+1} + g \sigma_z^i \Big] \,,
\label{eq:Hising}
\end{equation}
where $\sigma_x,\sigma_z$ are the Pauli matrices, and  the three-state Potts model
\begin{equation}
{{H}_{Potts}}(g) = -  \sum_i  \Big[ \Big( \sigma_i \sigma^\dagger_{i+1} +  \sigma^\dagger_i \sigma_{i+1} \Big) + g (\tau_i + \tau^\dagger_i) \Big] \,,
\label{eq:Hpotts}
\end{equation}
with the matrices $ \sigma = \sum_{s=0,1,2} \omega^s \ket{s}\bra{s}$, $\omega = e^{i2\pi/3}$ and $\tau =  \sum_{s=0,1,2} \ket{s}\bra{s+1}$, where the addition is modulo 3.
In both cases, the critical point ($g=1$) is described by a CFT. We consider
 different tMPO's lengths $T$, which acts as IR cutoff, investigating also the dependence of the results on  $\beta_0$ having the role of a UV cutoff.
 In order to make more connections with the CFT, we will consider both free and fixed boundary conditions (BC). In our formulation, the former corresponds to a $\ket{\uparrow}$ state for Ising and a $\ket{111}/\sqrt{3}$ state for Potts, while the latter is a $\ket{+}$ state in Ising and $\ket{001}$ in Potts \cite{tang2017}.
 Our calculations are performed using a Trotter step of  $\delta t=0.1$.

\begin{figure}
  \includegraphics[width=.48\columnwidth]{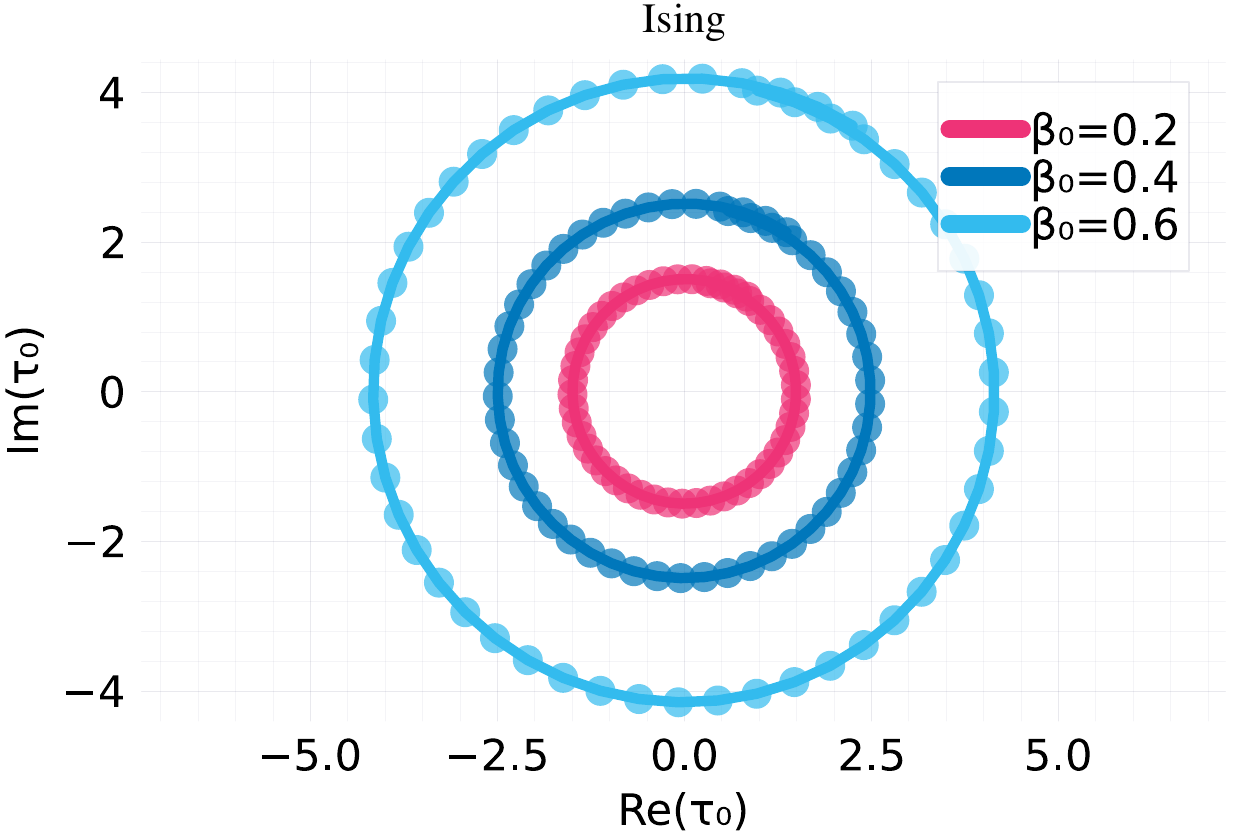}
    \includegraphics[width=.48\columnwidth]{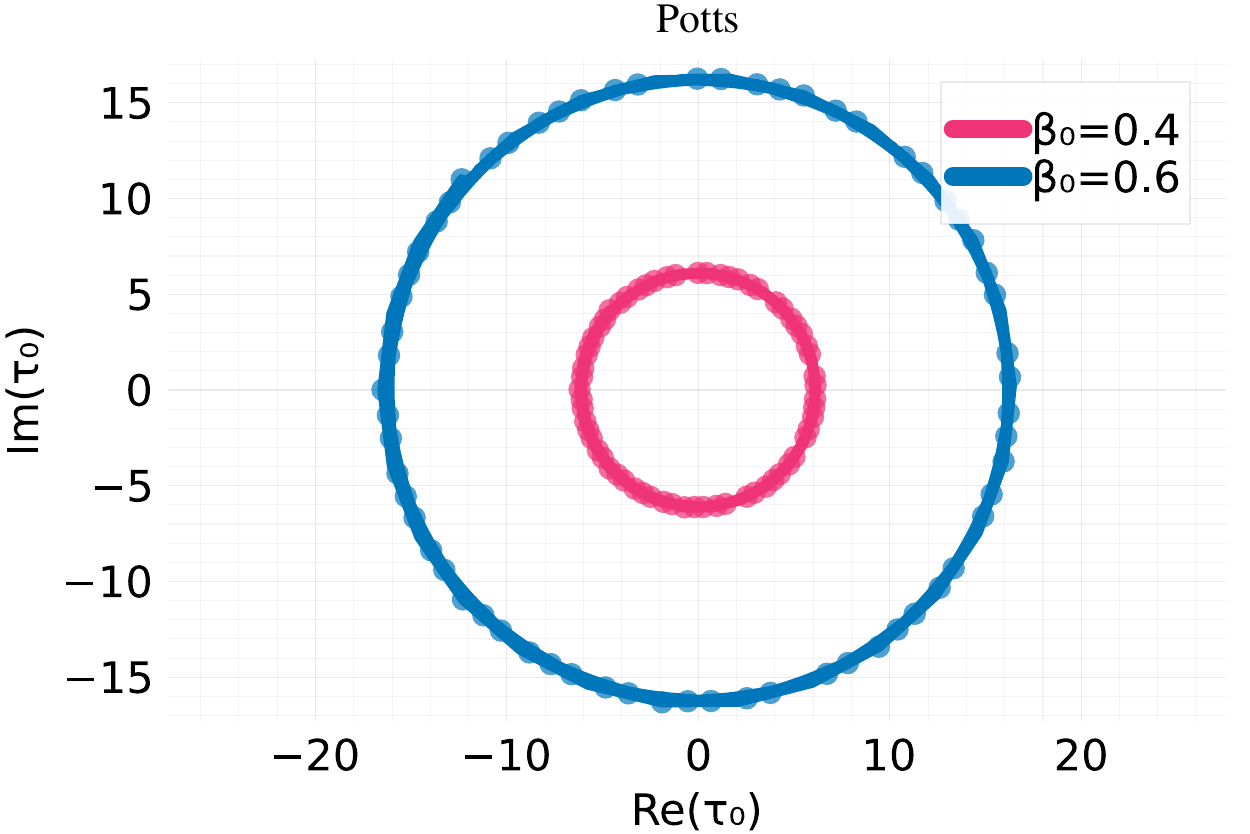}
 \caption{Real vs imaginary parts of $\tau_0$, the dominant eigenvalue  of $\Ts$ for various values of $T$ and $\beta_0$ for the Ising (left) and Potts (right) models. For a fixed $\beta_0$, as we vary $T$ they distribute on an almost constant radius circle.
 \label{fig:tau0s}}
\end{figure}

\subsection{Transfer matrix spectra.}
 As we vary the length $T$ of the tMPO $\Ts$ with $\beta_0$ fixed, its dominant eigenvalues become distributed over circles (see Fig.~\ref{fig:tau0s}), whose radius quickly collapses to a constant as $T$ increases, consistent with the predictions of Eq. \eqref{eq:lambda0exp}. We obtain the value of  $\kappa$ in the same equation from which we extract $c$ by fitting the numerical results for   $Im(\lambda_0)/T$ with a functional  form  $f_1(T) = a_0 + \kappa/T^2 + a_4/T^4$.

%

\begin{figure}
  \includegraphics[width=.48\columnwidth]{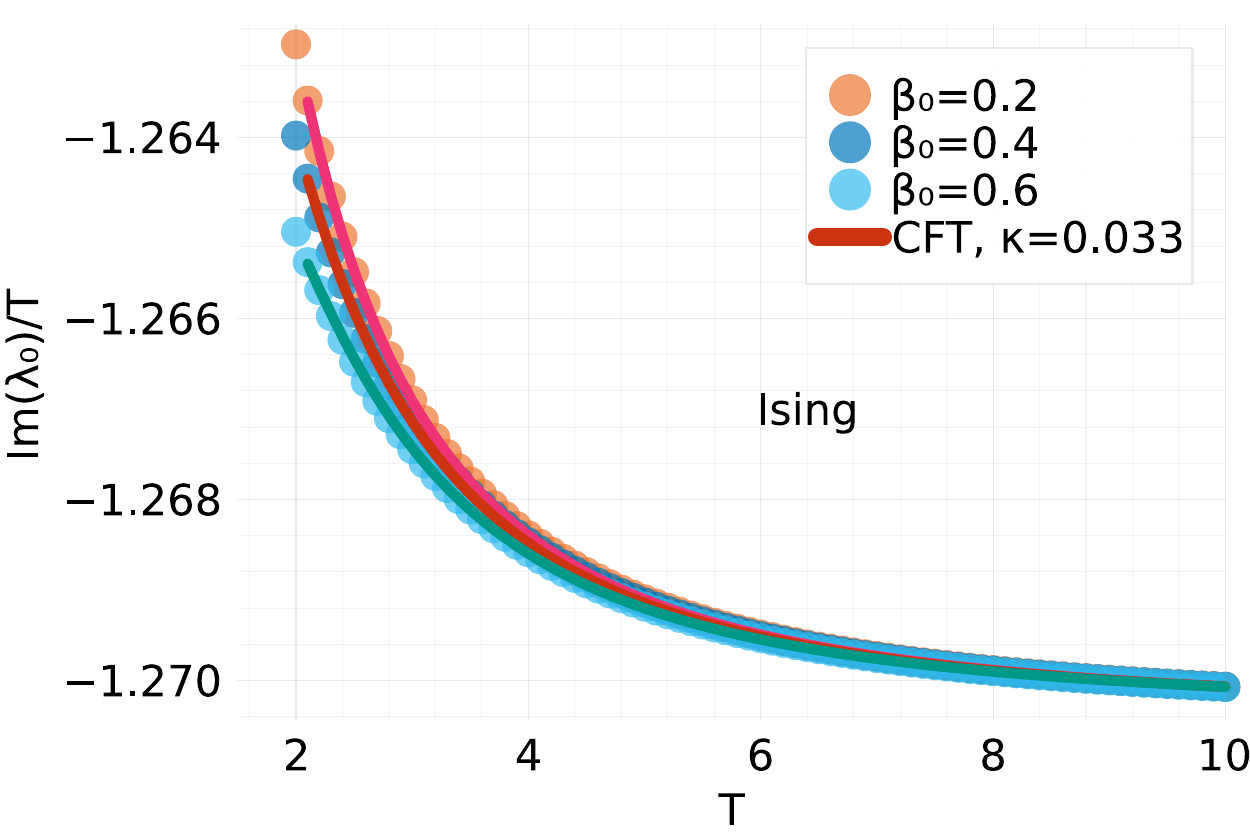}
  \includegraphics[width=.48\columnwidth]{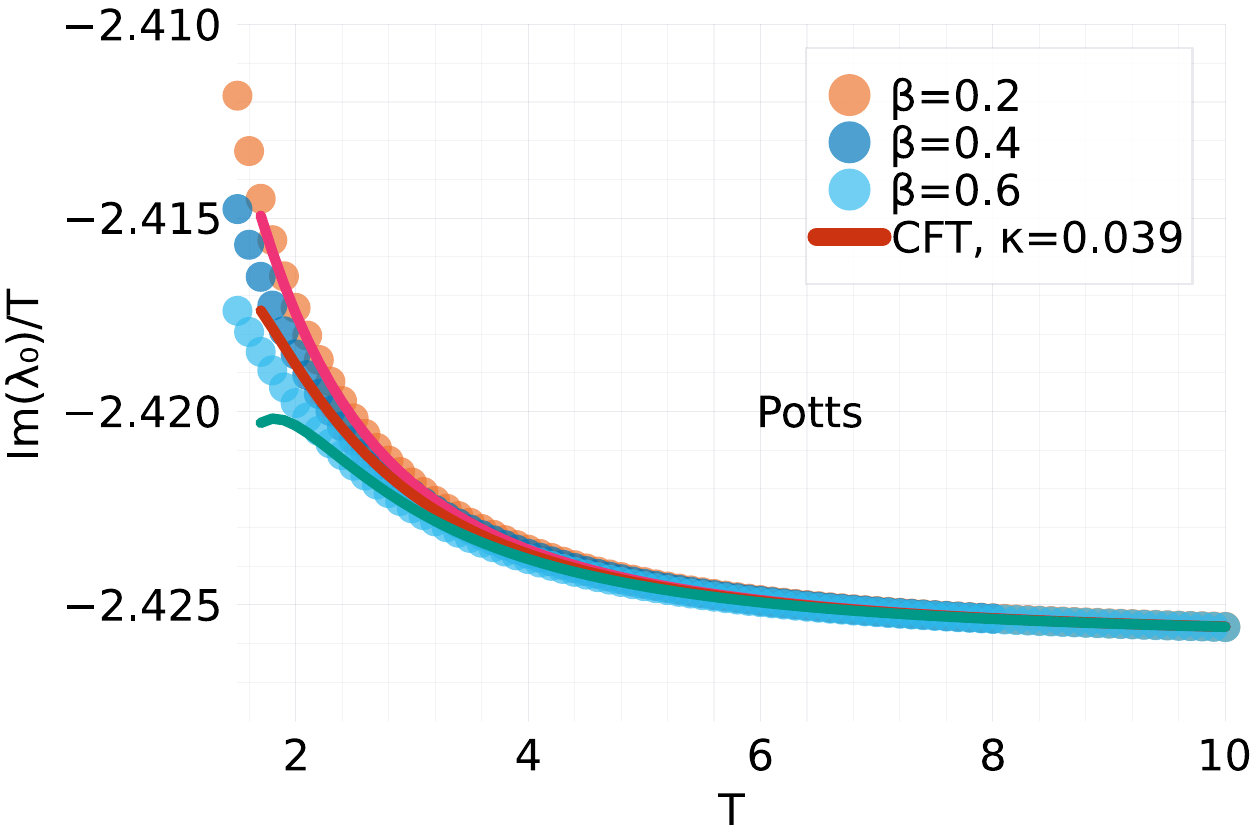}
  \includegraphics[width=.48\columnwidth]{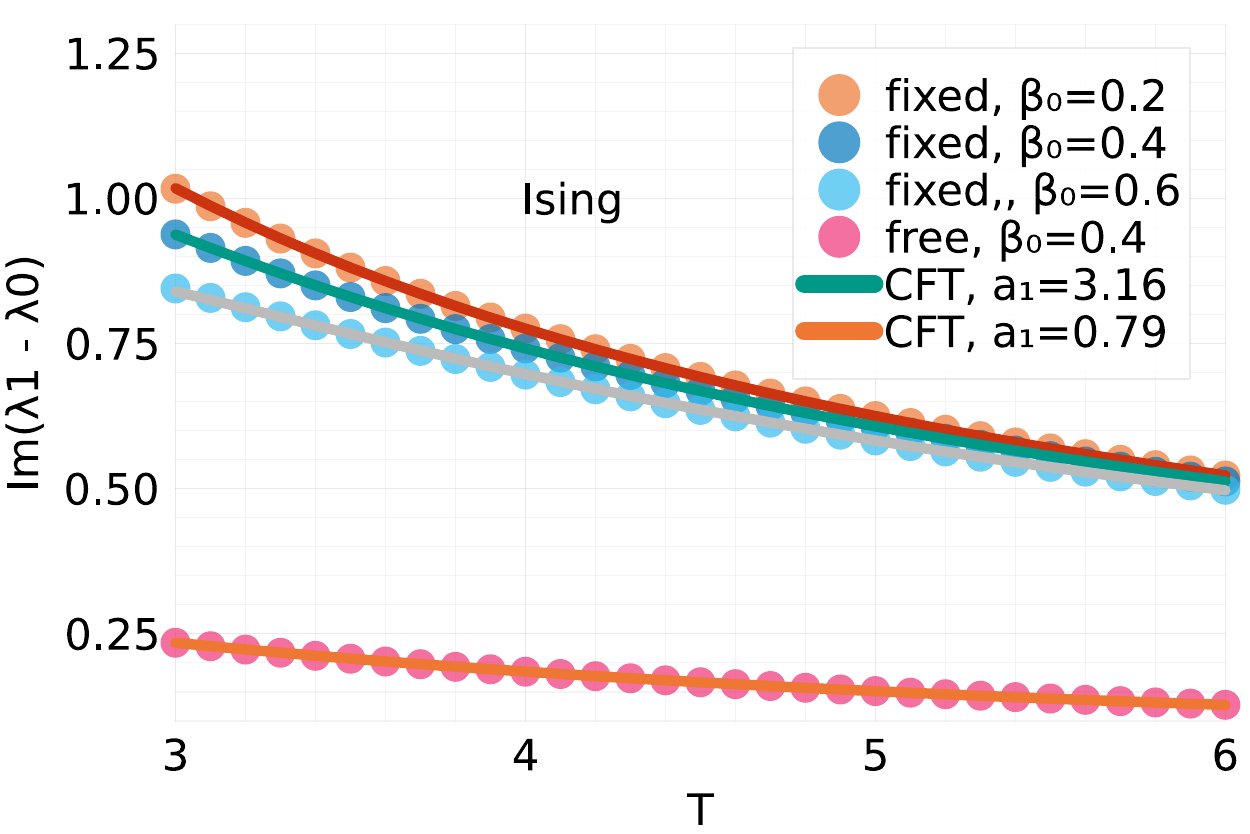}
    \includegraphics[width=.48\columnwidth]{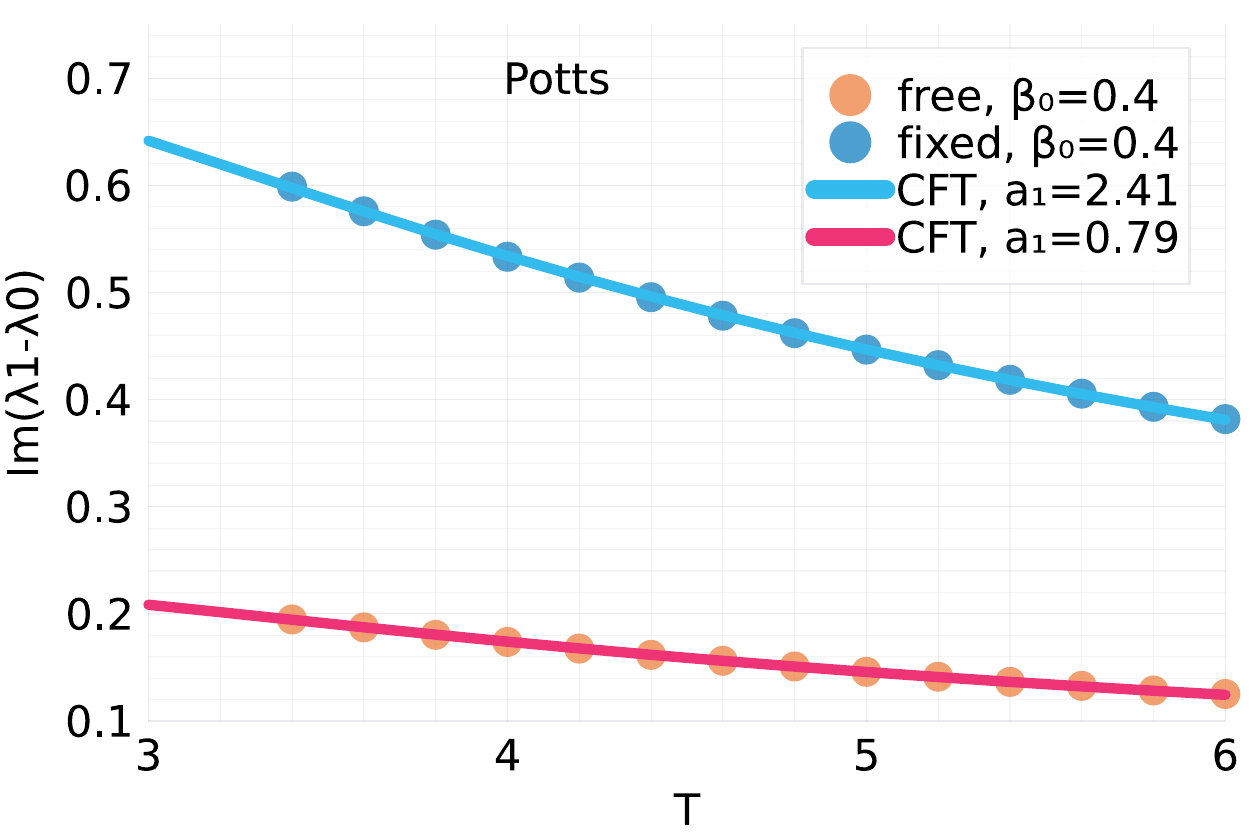}
 \caption{
 Top row: Imaginary parts of $\lambda_0$ for different values of $\beta_0$. The points are data from our TN calculation, solid lines are fits using the expected CFT form,
  which we use to extract the value of the central charge (see main text).  The fits show an excellent agreement with data for larger $T$, as expected.
 Bottom row: Imaginary part of the first gap $\lambda_1 - \lambda_0$ for different $\beta_0$ and BC, and corresponding fits to the CFT formulas to extract $x_1$.
  Left: Results for the Ising model, Right: Potts model.
  %
%
 \label{fig:im1_2}}
\end{figure}

The results for different values of $\beta_0$ are shown in Fig.~\ref{fig:im1_2} (top). All of them give comparable results, for Ising we obtain
$a_0 \simeq 1.27$ and $\kappa \simeq 0.032$, which from \Eq{eq:lambda0exp} with  $v=v_{Ising}=2$ \cite{tirrito2022}, gives $c_{fit}=0.49$, in excellent agreement with $c_{Ising} = 1/2$.
For Potts, the fits give $\kappa \simeq 0.039$ and using for the velocity $v_{Potts} \simeq 3\sqrt{3}/2$ \cite{eberharter2023} we obtain a value $c_{fit} \simeq  0.78$, compatible with the expected $c_{Potts} = 4/5$.
A similar analysis for the real part of $\lambda_0$ (see Appendix D) also confirms the  CFT predictions.

From the first excitation $\lambda_1$ of $\Ts$ we extract the first gap. Focusing again on the imaginary parts, we fit $Im(\lambda_1-\lambda_0)$ with
$f_3(T) = a_1/T + a_3/T^3$, inspired by \Eq{eq:gapsexp}, see Fig.~\ref{fig:im1_2} (bottom).
The results also confirm the CFT predictions \cite{cardy1986a}:
For the Ising model with fixed BC we obtain $a_1 \simeq 3.155$, in agreement with $\pi x_1/v_{Ising}$ for $x_1 = 2$.
For free BC we get instead
 $a_1=0.786 \simeq \pi/4$, consistent with  $x_1 = 1/2$.
 %
 %
%
%
For Potts,
we obtain $a_1 \simeq 0.788$ for free BC and $a_1 \simeq 2.41$ for fixed BC state, respectively.
This corresponds to
$x_1 = v_{Potts} a_1 / \pi \simeq 0.65$ and $1.98$, which we can match with the expected $2/3$ for the free and $2$ for  fixed boundary conditions \cite{cardy1986a,affleck1998}.

\subsection{Generalized entropies.} 
\begin{figure}
 \includegraphics[width=\columnwidth]{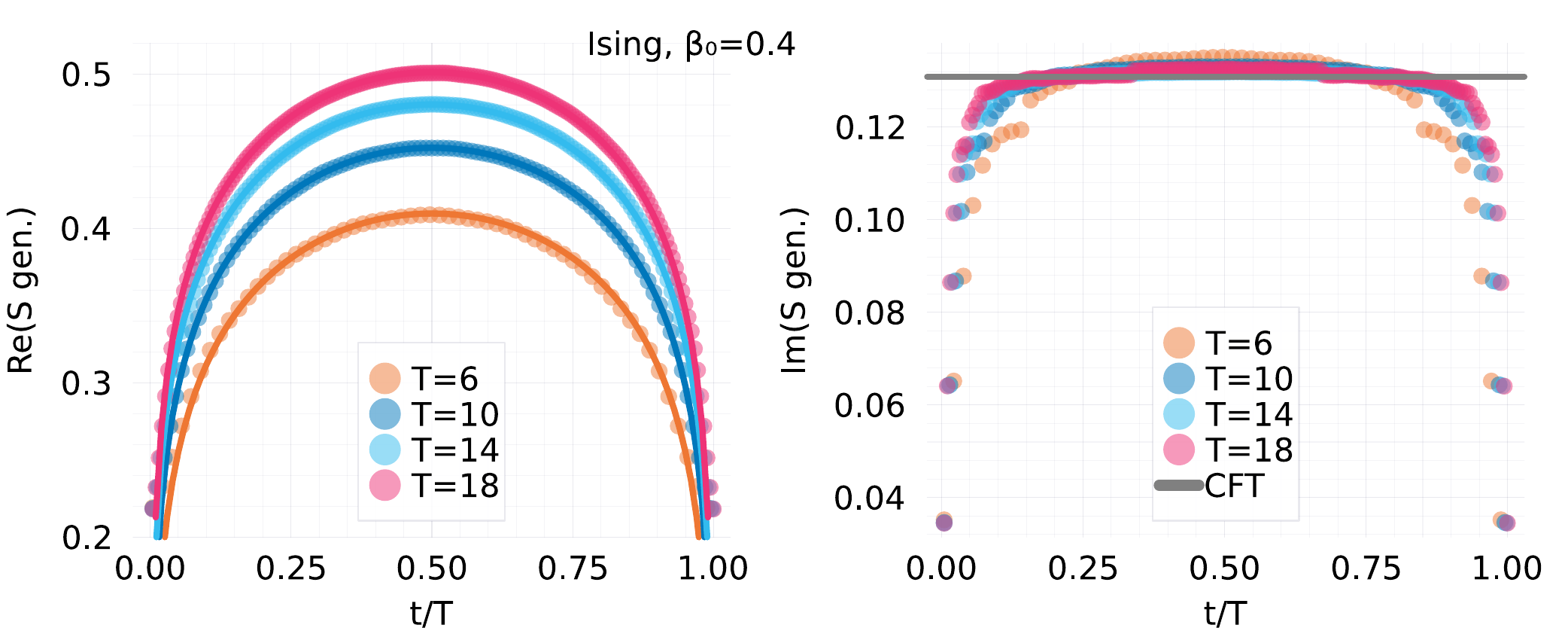}
  \includegraphics[width=\columnwidth]{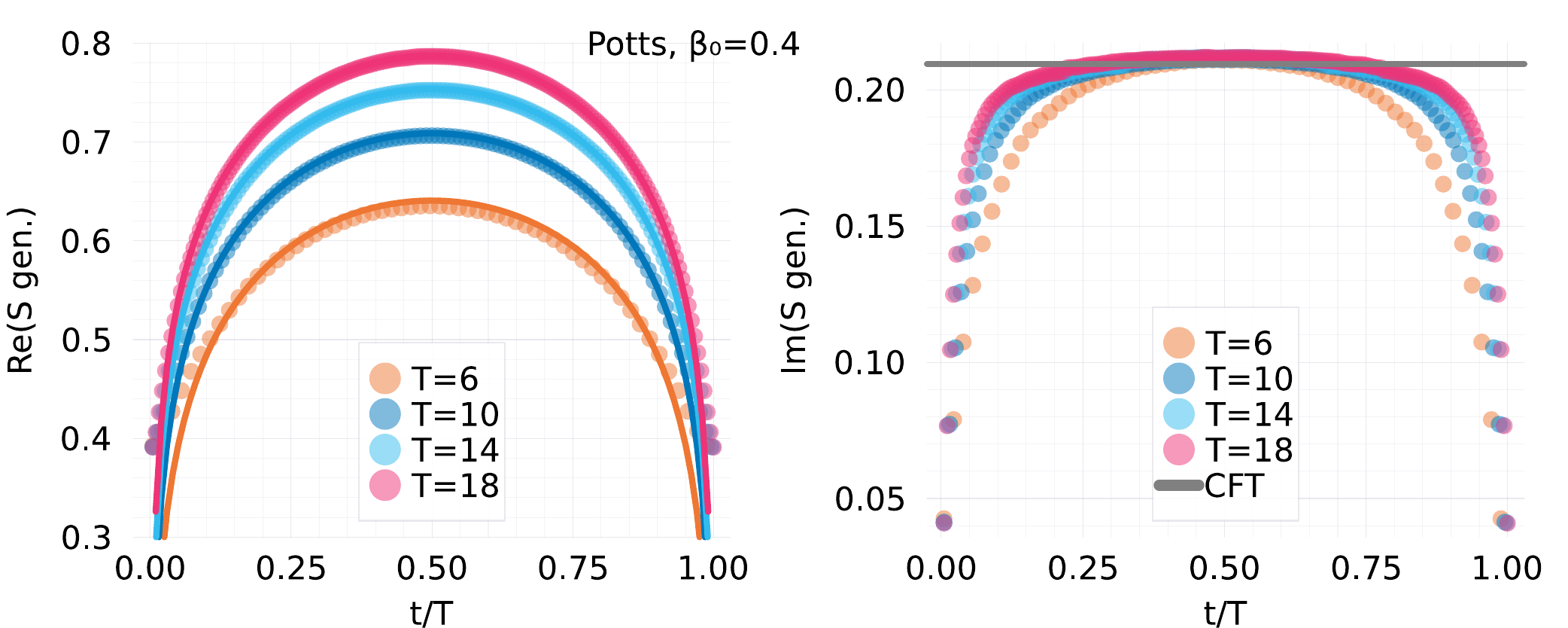}

 \caption{Numerical results for the real (left) and imaginary (right) parts of the generalized entropies for the critical Ising
 (top row)
 and critical Potts
  (bottom row) {with free BC}.
 Solid lines denote the CFT predictions \eqref{eq:Sgen_cft}. For Ising, we use $c=1/2$ central charge and $s_0 = 0.3$ the constant providing the best match for the real part,
 whereas for Potts we use the central charge $c=4/5$ and $s_0= 0.46$.
 The agreement with the CFT predictions is excellent for both real and imaginary parts, and it improves as we go to longer chains, as expected.
 \label{fig:genents}}
\end{figure}

The (complex-valued) generalized entropies are obtained from the dominant eigenvectors of the TM. They have an holographic meaning \cite{narayan2015,narayan2016,nakata2021,doi2023,doi2023a,narayan2023,li2023,shinmyo2023}, and  provide a measure for the complexity of the TN contraction and thus for the cost for simulating the \echo with tensor networks \cite{carignano2024b,banuls2009,lerose2023,hastings2015}.
 The results
 in \Fig{fig:genents} show that {both for Ising and Potts} the agreement with the CFT predictions is excellent already for relatively short $T$. Neglecting the boundary effects, the imaginary part of the entropy becomes constant and matches the expected $Im(S_{\rm {CFT}} )= \pi c/{12}$,
 and the real part is well fitted by \Eq{eq:Sgen_cft}.

 Different boundary conditions can affect the rate of convergence of the curves to the CFT predictions.
 The spectrum of $\mathcal{T}$ is that of a boundary CFT, so the operator content depends on the boundary conditions \cite{cardy1986a}.  Similarly, the entanglement spectrum is mapped to a different bCFT spectrum.
 This boundary spectrum gives rise to unusual finite size corrections of the type $T^{-x_n}$, which depend both on the Renyi entropy index $n$ and the boundary spectrum in the form $x_n=x_\alpha/n$ \cite{cardy2010,cho2017}. Here, the dependence on the boundary conditions is in the allowed values of $x_\alpha$. The detailed numerical analysis of the consequences of these observations are presented in the Appendix.

Our numerical results also validate equivalent predictions for the generalized entropies which have been obtained with different techniques based on holography and CFTs in \cite{nakata2021,doi2023a,guo2022,guo2024}.

\begin{figure}
  \includegraphics[width=.48\columnwidth]{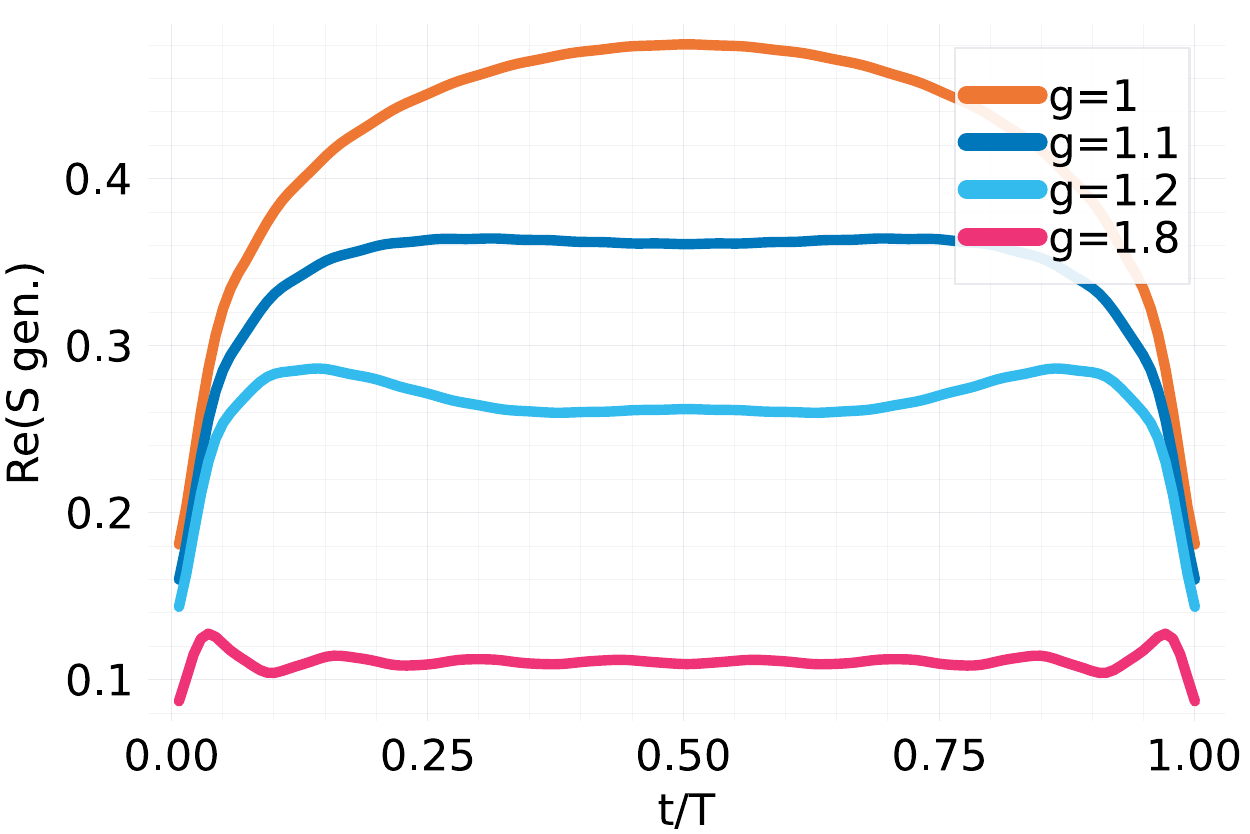}
 \includegraphics[width=.48\columnwidth]{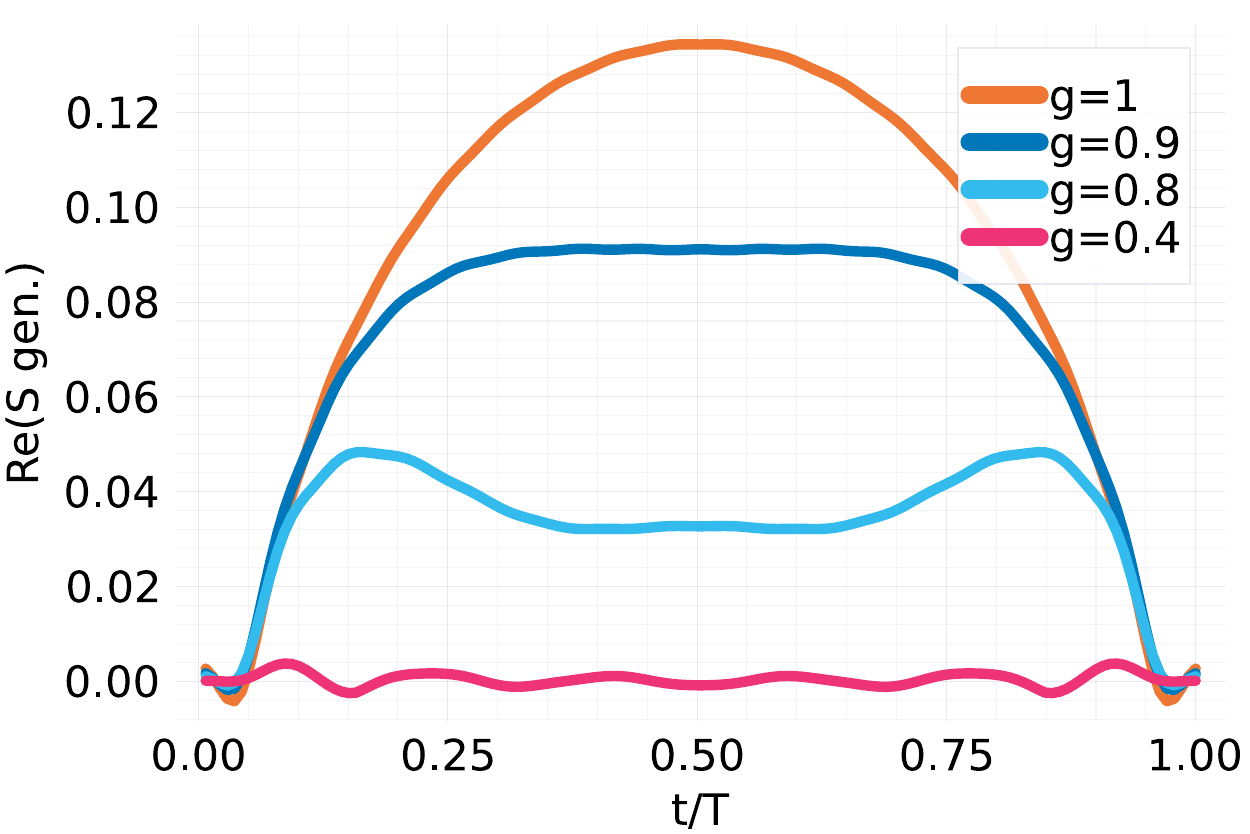}
 \caption{Real part of the generalized temporal entropies for the dominant vectors of the transfer matrix of the Ising model for $T=14$ and $\beta_0 = 0.2$. As we move away from the critical point $g=1$, the entropies saturate, exhibiting an area law. The left plot corresponds to fixed boundary conditions, the right to free ones.
 \label{fig:arealaw}}
\end{figure}

We finally check numerically in \Fig{fig:arealaw} our prediction of an area law for the temporal entropies as we move away from criticality. When moving away from the critical point the generalized temporal entropies exhibit the same behavior at the boundaries of the temporal chain but saturate in the middle. This generally results in lower bond dimensions required for the description of the dominant vectors in terms of tMPS, opening the possibility for efficient simulations of out-of-equilibrium dynamics.

\section{Conclusions}

In this work, we have unveiled the universal behavior of the \echo after a quench to a critical Hamiltonian. We show that its leading decay is dictated by the central charge of the underlying CFT, while finite-time corrections are governed by the critical exponents of the theory.
By leveraging temporal matrix product states, we find that the computational complexity of calculating the \echo is controlled by the growth of temporal entropies. Using CFT, we derive a closed-form expression for the scaling of these temporal entropies as a function of time, demonstrating a logarithmic growth. Consequently, the complexity of simulating the \echo with MPS reduces from exponential to polynomial in time. This result establishes that the computation of the \echo is feasible with classical algorithms (for alternative approaches that similarly reduce simulation complexity, see \cite{surace2019a,strathearn2018,white2018,rakovszky2020,frias-perez2022,liu2024a}).
Thus, we can perform accurate classical simulations of this scenario using tensor networks, confirming our analytical predictions.
Moreover, our numerical simulations extract the key universal properties, supporting the idea that these can also be probed experimentally by studying the \echo for relatively short times.
Finally, our results reveal that the time evolution operator generated by a critical Hamiltonian leads, at long times, to a unitary transfer matrix in space—an outcome expected in dual-unitary dynamics. Conformal symmetry thus seems to naturally lead to the emergence of dual-unitary behavior at large times.

\begin{acknowledgments}
We would like to thank M.C. Bañuls, P. Calabrese, E. Lopez,  G. Sierra, E. Tonni for discussions related to this project. We thank A. Tomut for suggesting the reference \cite{tang2017} clarifying the role of the lattice BC for the Potts model.
LT acknowledges support from the Proyecto Sinérgico CAM Y2020/TCS-6545 NanoQuCo-CM,
the CSIC Research Platform on Quantum Technologies PTI-001, and from the Grant TED2021-130552B-C22 funded by MCIN/AEI/10.13039/501100011033 and by the ``European Union NextGenerationEU/PRTR'',
and Grant PID2021-127968NB-I00 funded by MCIN/AEI/10.13039/501100011033.
\end{acknowledgments}

\bibliographystyle{quantum}
\bibliography{loschbib}


\onecolumn
\begin{appendix}

\section{Mapping the quench geometry to CFT}
At the core of this work lies the mapping of the amplitude associated with the return probability to the initial state  after the evolution for a time $T$ after a quench to the critical point of a lattice Hamiltonian with the path integral of a field theory on a specific geometry which can be studied using conformal field theory.

This mapping greatly simplifies given that the Hamiltonian is defined at a critical point where, due to universality, strong fluctuations are significant, though their microscopic details are irrelevant in the renormalization group sense. By employing a formalism that links thermal and quantum fluctuations in statistical mechanics and quantum field theory, we can translate these analogies into computational tools for predicting physical phenomena.

The idea is that the partition function of a statistical system with short-range interactions can then be seen equivalently as either a sum over classical variables in a d-dimensional euclidean space with a classical Hamiltonian $H({s_i})$, or as the trace of a time evolution operator
$U(\tau) = e^{-\tau \hat{H}(\{{\phi\})}}$
 associated with a quantum Hamiltonian $\hat{H}$ in $d-1$ dimensions of certain appropriate variables $\phi$.
 Furthermore, for our problem we can equivalently adopt either an operator or a functional approach, which allows to substitute the sum over the classical discrete variables ${s_i}$ in terms of a path integral on continuous variables $\phi(x)$, weighted by the classical action $S$.  We thus have

\begin{equation*}\begin{alignedat}{2}
Z & = \sum_{\set s} e^{-H({{s_i}})} \quad & = \quad {\rm Tr}_\phi \prod_\tau e^{-\tau \hat{H}({\phi})} & \approx \quad \int {\mathcal{D}}\phi e^{-S[\phi]} \,, \\
   &  {\text {stat mech}}                             &  \quad \text{quantum time evol}                                     &  \qquad \quad \text{QFT} \nonumber
\end{alignedat}\end{equation*}

and we can interpret our setup either as a Trotterized time evolution in an appropriate basis or a field theory discretized on a lattice with a properly defined continuum limit. Such mapping is
particularly simple at a critical point, when the correlation length is much larger than the lattice spacing and using universality arguments we can focus on very simple lattice models.

In the following, we recall some of the ideas revolving around the equivalence of these formalisms, referring the interested reader to the standard literature \cite{kogut1979,cardy1988,francesco1997,henkel1999,cardy2008} for additional details (see also \cite{okunishi2022} for a more historical overview related to the development of tensor network algorithms). 

Our starting point for the computation of the \echo is the amplitude $\bra{\psi_0}e^{-i\hat{H}T}\ket{\psi_0}$, where $\hat{H}$ is the quantum Hamiltonian of our system.
In the Feynman path integral formalism, this amplitude can be mapped to a functional integral over the action $S$. 
The derivation for a zero-dimensional problem (ie. a single particle) is textbook material: one begins by splitting the time evolution operator $U(T) = e^{-i\hat{H}T}$ in smaller intervals $ T = N_T \delta T$, an operation which will allow to work with a Trotter approximation for non-commuting terms in the exponential. More specifically,
taking a generic nonrelativistic Hamiltonian $\hat{H} = \hat{K} + \hat{V}$, $\hat{K}$ being the kinetic term and $\hat{V}$ the potential, we consider an infinitesimal time evolution operator sandwiched between two position eigenstates 
\beq
\braket{x'|U(\delta T)|x}  = \braket{x'|e^{-i\hat{K}\delta T}e^{-i\hat{V}\delta T} + {\mathcal{O}}(\delta T)^2 )|x}  \sim \exp\Big\{ i \delta T \Big[ \frac{m}{2} \frac{(x-x')^2}{\delta T^2} - V(x) \Big] \Big\}  \,,
\label{eq:propagator_action}
\eeq
where we inserted a complete set of momentum eigenstates and performed the momentum integration, neglecting terms of order $\delta T^2$. 
We can then identify the result as the exponential of the action $S$ for this infinitesimal step from $x$ to $x'$, and putting together the various amplitudes we arrive at 
$\braket{x_f|U(T)|x_i} = \int {\mathcal{D}}x e^{iS(x)} $,
where the integral  $\int {\mathcal{D}}x$ is meant as a sum over all possible intermediate positions of the particle, that is, a sum over all paths.

At this point, it is convenient to perform the standard analytic continuation to imaginary time $t \rightarrow -i \tau$, $\tau \in [0,\beta]$, which also allows to make 
a connection with statistical mechanics. In particular, the path integral expression for the transition amplitude becomes equivalent to a classical partition function
for a one-dimensional system, whose sites are labelled by the different time steps of the time evolution. The (now euclidean) action describes the coupling 
of the position variables via the discretized derivative (cf.~\Eq{eq:propagator_action}) for different time steps. 
In turn, this implies that we can interpret the infinitesimal propagator $\braket{x'|U(\delta \tau)|x}$ as an element ${\mathcal{T}}(x',x)$ of the transfer matrix  evolving the system 
from the position $x$ to $x'$ at the next timestep, and the whole amplitude can be written as
\beq
Z = \int \prod_i dx_i {\mathcal{T}}(x_{i+1},x_i) = \int \braket{x_f | {\hat{\mathcal{T}}}^{N_T} | x_i} \,,
\eeq
so that the transfer matrix contains all the information about the evolution of the system. We can then use the transfer matrix built upon the time evolution operator as a bridge between 
the path integral to the Hamiltonian formalism, a connection which becomes clear in the limit of infinitesimal timesteps, where one has $\hat{{\mathcal{T}}} \approx 1 - i \delta\tau \hat{H} $.

The same type of considerations can be naturally extended to higher-dimensional systems, simply interpreting the transfer matrix as a linear operator that builds the partition function of a $d$-dimensional statistical lattice model with short-range interactions by relating the states
of two adjacent $(d-1)$-dimensional slices. 

More specifically, for the dynamics of a one-dimensional spin chain such as the ones we consider in this work, we can map our return amplitude after imaginary time rotation to a two-dimensional classical statistical model. 
Since one of the two dimensions will be associated with euclidean time, we can define the classical model on an anisotropic grid with different couplings in the two directions, in order to allow for a separate treatment of the two directions.
We now consider two rows of spins aligned along the spatial direction and define a generalization of the transfer matrix from above, which now connects the two rows,
and has a number of constituents which grows exponentially with the system size. This transfer matrix again provides the link between the formalisms, as it can both be used to construct the full 2d partition function and to connect
to the underlying quantum Hamiltonian in the limit of vanishing timesteps, after appropriately correcting the statistical model couplings to describe the same physics in such limit.

An analogous mapping for this setup can be done between the two-dimensional statistical mechanics problem and quantum field theory in 1+1 (time+space) dimensions after imaginary time rotation.
 Here the analogy is even more transparent, as quantum field theory naturally incorporates time, and if we regularize its path integral on a grid and dissect it along timelike slices, we end up with the same structure as the one discussed above, allowing the connection with the Hamiltonian formalism. 

At a critical point the continuum limit is well-defined, allowing us to disregard microscopic details and focus on the long-wavelength properties of the time-evolved one-dimensional quantum system or the equivalent 2d classical statistical system. These properties can be characterized using a corresponding quantum field theory in the same universality class. Importantly, at criticality, this field theory is massless and conformal, enabling us to utilize the extensive tools developed for characterizing CFTs in 1+1 dimensions.

As a result, we can obtain from CFT predictions for the quantities of interest, the transition matrix ${\mathcal{T}}$ and the Loschmidt echo ${\mathcal{L}}$.
 The latter can thus be seen as equivalent to a partition function, and its intensive part $l$ is finite and plays the role of a free energy in statistical mechanics. 

We finally note that all these equivalences arise in a natural way in the pictorial description of tensor networks: Fig. 1(c) in the main text can be seen equivalently as the partition function of a two-dimensional lattice system, as a Trotterized time evolution of a one-dimensional chain, as well as a functional integral which is UV-regularized by discretization on a lattice.

\section{CFT Predictions for the Transfer Matrix} 

Having shown how the return amplitude for a one-dimensional chain can be mapped to a path integral in 1+1-dimensional CFT, let us now recap some of the main ideas behind the derivations for the transfer matrix spectrum which we employ in our manuscript. For more details, we refer to the reviews \cite{cardy1988,ginsparg1988,cardy2008}.

A conformal field theory (CFT) is a field theory whose action has a special symmetry since it is invariant under  conformal transformations. Such a symmetry is typically expected to be a consequence of scale invariance and locality of the Hamiltonian. As a result, the physics of many critical points of lattice models can be described at large distances by a CFT.
From the action $S$ of the field theory, one can define the stress energy tensor $T_{\mu\nu}$ as the response of the action to an infinitesimal change of coordinates, $r^\mu\to r^\mu+\alpha^\mu(r)$,
\begin{equation}
 \delta S = - \frac{1}{2\pi}\int \left(T_{\mu \nu} (r) \partial^{\nu}\alpha^\mu(r) \right)dr,
\end{equation}
and as such the stress tensor can be viewed as the generator of scale and conformal transformations.
In 2D the conformal algebra becomes infinite dimensional and the CFT is highly constrained. In particular, any analytic change of coordinates becomes a symmetry of the theory, and thus rather than thinking of the CFT as defined on a 2D space-time, one usually thinks of it as defined on the complex plane. 
 It is worth noting already at this point that in Euclidean CFT the two coordinates play the same role, an aspect which will become important for us when we later focus on a transverse transfer matrix.
 If we consider the field coordinates  $z \in {\mathbb{C} }$ and its conjugate $\bz$, we can  map the theory from the plane to other geometries via any analytic changes of coordinates. 
%

Scale invariance is enough to understand that correlation functions in a CFT decay algebraically, e.g. for a two-point function on the infinite plane we have
\begin{equation}
\braket{\phi(z_1) \phi(z_2)}=\frac{1}{(z_1-z_2)^{2h}(\bz_1-\bz_2)^{2\bh}} \,,
\end{equation}
dictated by a set of critical exponents $\set{h, \bar{h}}$ that define the universality class of the model.

Conformal symmetry now dictates that, under a coordinate change $\omega = f(z)$,
\begin{equation}
\braket{\phi(\omega_1) \phi(\omega_2)} = 
\frac{\braket{\phi(z_1) \phi(z_2)}}{(f'(z_1))^h(f'(z_2))^h(f'(\bz_1))^\bh(f'(\bz_2))^\bh} \,,
\end{equation} 
and we now just need to express the old coordinates in terms of the new ones.

For example, if we want to describe the physics on a cylinder rather than a plane, 
as in the case of thermal states or ground states on finite rings, 
we can define $\omega~=~\frac{\beta}{2\pi}\log(z)$, where $\omega = s + i u$ is now defined on the desired cylinder with circumference $\beta$ along the imaginary axis.
We then have $f'(z) = \frac{\beta}{2\pi z}$ and $z(\omega) = \exp(2\pi\omega/\beta)$,
and the two point function on the cylinder becomes
\begin{equation}
\braket{\phi(\omega_1) \phi(\omega_2)} = \frac{ \left(\pi/\beta\right)^{2x}}{\sinh\Big(\frac{\pi}{\beta}(\omega_1-\omega_2)\Big)^{2h} \sinh\Big(\frac{\pi}{\beta}(\bw_1-\bw_2)\Big)^{2\bh}} \,,
\label{eq:correl_omega}
\end{equation}
with the exponent $x = h + \bh$. 

Conformal symmetry also dictates how the stress energy tensor changes under a conformal map,
\begin{equation}
 T(\omega)\to T(z) = f'(\omega(z))^2 T(z)+\frac{c}{12}\{z,\omega\} \,,
 \label{eq:se_te}
\end{equation}
with $\{z,\omega\}$ the Schwartzian derivative of $f$.

In the traditional Hilbert space formalism,
the quantum Hamiltonian $\hat{H}$ is related to the integral of the time-time component of the stress-energy tensor ${\mathbb T}_{\mu\nu}$ over the space-like curve on which one quantizes the theory\footnote{This is nothing but the euclidean version of the QFT statement that the Hamiltonian is the space integral of the time-time component of the energy-momentum tensor.}.

It is natural  to associate the imaginary axis of the complex coordinates with the temporal direction. We can then write
\begin{equation}
 \hat{H} =  \frac{1}{2\pi} \int ds {\mathbb T}_{uu}(s) \,.
 \end{equation}
  As seen in the previous section, the quantum Hamiltonian is intimately related to the transfer matrix, which is the object we are interested in for characterizing our system.

Using the rule that dictates how the stress energy tensor changes under  conformal maps in \Eq{eq:se_te}, we can thus predict the form of the transfer matrix in different geometries.
For example, the spatial transfer matrix defined on an infinitely long cylindrical geometry with radius $\beta$ reads
\begin{align}
\Ts & = \exp{\left[- \Bigg(\frac{\kappa}{\beta}+\frac{2\pi}{\beta}(L_0+\bar{L}_0) \Bigg)\right]}\,, 
\label{eq:tm_cyl2}
\end{align}
where ${\kappa}= -{\pi c \delta t}/{6}$, $c$ being the central charge.
Here $ L_0 = \frac{1}{2\pi i} \oint dz z {\mathbb{T}}(z)  $ is the generator of the holomorphic part of Virasoro algebra in the plane and $\bar{L_0}$ of the anti-holomorphic part of it, and ${\mathbb{T}}(z) = {\mathbb{T}}_{z z}$ in the original
 complex plane coordinates.
 This in turn implies that the eigenvalues of $\hat{H}$ are in one to one correspondence with those of $L_0$, and thus the scaling operators of the theory, which will depend on the corresponding scaling exponents $x$. These results thus allow us to completely determine the spectrum of the transfer matrices in our geometry.

As discussed in the main text, the \echo of a product state requires working with boundaries representing the initial and final states, resulting in a strip-like geometry (see Fig.1 main text).
This can be obtained via the mapping\footnote{Here we strictly focus on CFT results and omit additional factors which enter the calculations performed on the lattice model such as the sound velocity and the non-universal terms (see main text).}
  $\omega \equiv s +i u =\tfrac{\beta}{\pi}\log( z )$.

  The corresponding integral of the energy tensor then gives
  \beq
 {\hat H} = \frac{\pi}{\beta} L_0 - \frac{\pi c}{24 \beta} \,,
 \eeq
leading to the following expression for the spectrum of the transfer operator:
\begin{align}
\Ts & = \exp{\left[- \Bigg( \frac{\kappa_{s}}{\beta}+\frac{\pi}{\beta} L_0 +  \mathcal{ O}\Big(\frac{1}{\beta^2}\Big) \Bigg) \right]}\,, \label{eq:tm_cyl}
\end{align}
where now ${\kappa_{s}}= -{\pi c \delta t}/{24}$.

We can extract even more informations from the CFT. In particular,
if we consider \Eq{eq:correl_omega} when
the two points are separated only along the temporal direction, their distance being $i(u_1-u_2)$, we obtain
\begin{equation}
\braket{\phi(\omega_1) \phi(\omega_2)}=\left(\frac{\pi}{\beta}\right)^{2x}{\sin\left[\frac{ \pi}{\beta}\left( u_1-u_2\right)\right]}^{-2x} \,. \label{eq:ent_ent}
\end{equation}
and, as discussed in the main text, 
 for twist fields  where $x$  becomes $\Delta_n = \frac{c}{24}\left(n-1/n\right)$, such a prediction gives access to the Tsallis entropies of order $n$ \cite{srednicki1993,callan1994,vidal2003b,calabrese2004,caraglio2008}, and consequently to the generalized temporal entropies of our problem after analytic continuation $n\to 1$.
  
%
%


\section{Tensor Network setup, symmetric MPO}
\label{app:setup}

In this section we briefly describe the methods used in this work for building the MPO of the transition matrix $\Ts$ and extract its eigenvalues. Calculations have been performed using the \verb|ITransverse.jl| library \cite{ITransverse}, built on top of \verb|ITensors.jl| \cite{ITensor}.

The concrete algorithm was introduced in \cite{carignano2024b}  inspired on those presented in \cite{banuls2009,hastings2015}. It is basically a power method that iteratively applies the transfer matrix $\Ts$ to an initial MPS state and uses a low rank approximation of the reduced transition matrices in order to compress the MPS.

Our temporal MPO $\Ts$ is defined by contracting one column of the infinite tensor network shown in Fig.~2 in the main text, containing  $N_T+2N_\beta$ (with $N_\beta = {\beta_0}/{\delta t}$) tensors.
The basic ingredients for the MPO are the tensors $W_i(\delta t)$ associated with a given site and timestep, building up the time evolution operator $U(\delta t)$ (see Fig.~1 in the main text). 

Since we work with translationally invariant systems, there is 
no explicit space dependence, and all tensors have the same dependence on $\delta t$. The only inhomogeneities in the time direction are induced by the initial state $\ket{\psi_0}$, which we take as a product state. 
 
For a gapped $\Ts$,  the intensive \echo converges exponentially fast to $ l =- |\lambda_0|/T$ and in order to extract $\lambda_0$ we use an MPS ansatz for the dominant eigenvectors 
 $\ket{R}$ and $\bra{L}$ of $\Ts$.
 
 The roles of virtual and physical legs are interchanged if one performs a transverse contraction, as the temporal MPO  $\Ts$ is applied sideways to the boundary MPS which will result in the dominant left and right vectors. In this sense, an asymmetry in the virtual legs of the MPO tensors will result in a left dominant vector for $\Ts$ which is different from the right one, ie. $\ket{L} \neq \ket{R}$. Nevertheless, we were able to work with tensors which are symmetric on both the physical and virtual legs, so that this additional complication does not arise.   
 More specifically, for the Ising model we employ the compact representation proposed in \cite{pirvu2010} at second-order in the Trotter expansion, for which we obtain for the MPO tensors
 \beq
 W = 
 \begin{pmatrix}
 \cos(\delta t) (a \mathbf{1}+ b \sigma_z)&  \sqrt{i \sin(\delta t)\cos(\delta t)} \sigma_x \\
  \sqrt{i \sin(\delta t)\cos(\delta t)} \sigma_x & i \sin(\delta t)  (a \mathbf{1}+ b \sigma_z) 
 \end{pmatrix} \,,
 \eeq
 with $a=1-2\sin(g\delta t/2)^2$ and  $b=2i\sin(g\delta t/2)\cos(g\delta t/2)$.

The generalization of this type of construction unfortunately would not lead to a symmetric left-right MPO in the case of the Potts model, due to the presence of both $\sigma_i$ and $\sigma^\dagger_i$ terms (cfr. Eq.~(11) in the main text). Nevertheless, we are still able to extract a symmetric expression by considering the two-body operator $U_{i,i+1}(\delta t)$ and separating it via a symmetric SVD factorization. The resulting MPO tensors are again symmetric in both physical and virtual legs.

Being able to work in this symmetric gauge in which $\bar{\Ts}=\Ts^t$ (where the bar denotes complex conjugation),  
 we can thus focus on extracting $\ket{R}$ and then obtain $\bra{L} = \bra{\bar{R}}$ by a simple transposition. This 
  also significantly improves the numerical stability of our TN simulations (see eg. \cite{tang2023a} and references therein for a discussion on the challenges presented by non-hermitian eigenvalue problems).

Having obtained the form of the transfer matrix tensors, we build the transverse MPO for $\Ts$ by incorporating $N_\beta$ steps of imaginary time ($\delta t \rightarrow -i \delta\beta_0$) at the edges of the TM, where the real time evolution is performed in the center.

The determination of the dominant eigenvalues of $\Ts$ was performed via a symmetric power method algorithm which simply goes as follows: starting from an initial guess for the right dominant vector $\ket{R}$ in the form of MPS along the temporal direction, at each iteration $\Ts$ is applied to $\ket{R}$ and the resulting MPS is truncated by optimizing the overlap $\braket{\bar{R}|R}$, since, as already discussed, in our case we have $\bra{L} = \bra{\bar{R}}$. This is done by appropriately truncating on the singular values of the reduced transition matrices, Eq.(8) main text (see \cite{carignano2024b} for additional details). The iterations are repeated until convergence is reached, and the eigenvalues of the transition matrices are used to compute the entropies along the temporal chains.

\section{Fits on real parts of the dominant eigenvalues. }
\label{app:realfits}

By fitting the real part of the dominant eigenvalues of $\Ts$ we can get further confirmation of our CFT predictions, although the procedure is a bit more involved. 
When it comes to the real parts,  
by comparing with Eq.~(4) in the main text we see that
a single term $a_0 = 2\beta_0 a v + b$  does not allow us to resolve the pieces involved, but we can extract some additional information by looking at the difference of the real parts
for different values of $\beta_0$. Indeed, if we consider
$ \Delta_\beta \lambda_0 \equiv \lambda_0^{\beta_0\equiv\beta_1} - \lambda_0^{\beta_0\equiv\beta_2} $, we expect that
$Re(\Delta_\beta \lambda_0) = 2a (\beta_1 - \beta_2)  + \mathcal{ O}(1/T^2)$, whereas $Im(\Delta_\beta \lambda_0)=  \mathcal{ O}(1/T^3)$.

We thus fit the difference of the real part for different values of $\beta$ with a functional form $f_2 = b_0 + b_2/T^2$.
As an example, we fit for the Ising model,
for the case $\Delta_\beta\lambda_0(\beta=0.4 - \beta=0.2)$ we get 
$  b_0 = 0.51 $, whereas for  $\Delta(\beta=0.6 - \beta=0.2)$ we find  $b_0 = 1.02$, showing again an excellent agreement with the CFT prediction which indeed gives a ratio of 2 between the two values.

At this point,
we can also extract the non-universal coefficient $a = b_0/2 (\beta_1 - \beta_2)  \simeq 1.275 $ from here, which is consistent with the $a_0$ from the fit of the imaginary part above.

\section{Corrections to the leading finite size scaling for the generalized entropies}
\label{app:scaling}
\begin{figure}
\begin{center}
 \includegraphics[width=.6\columnwidth]{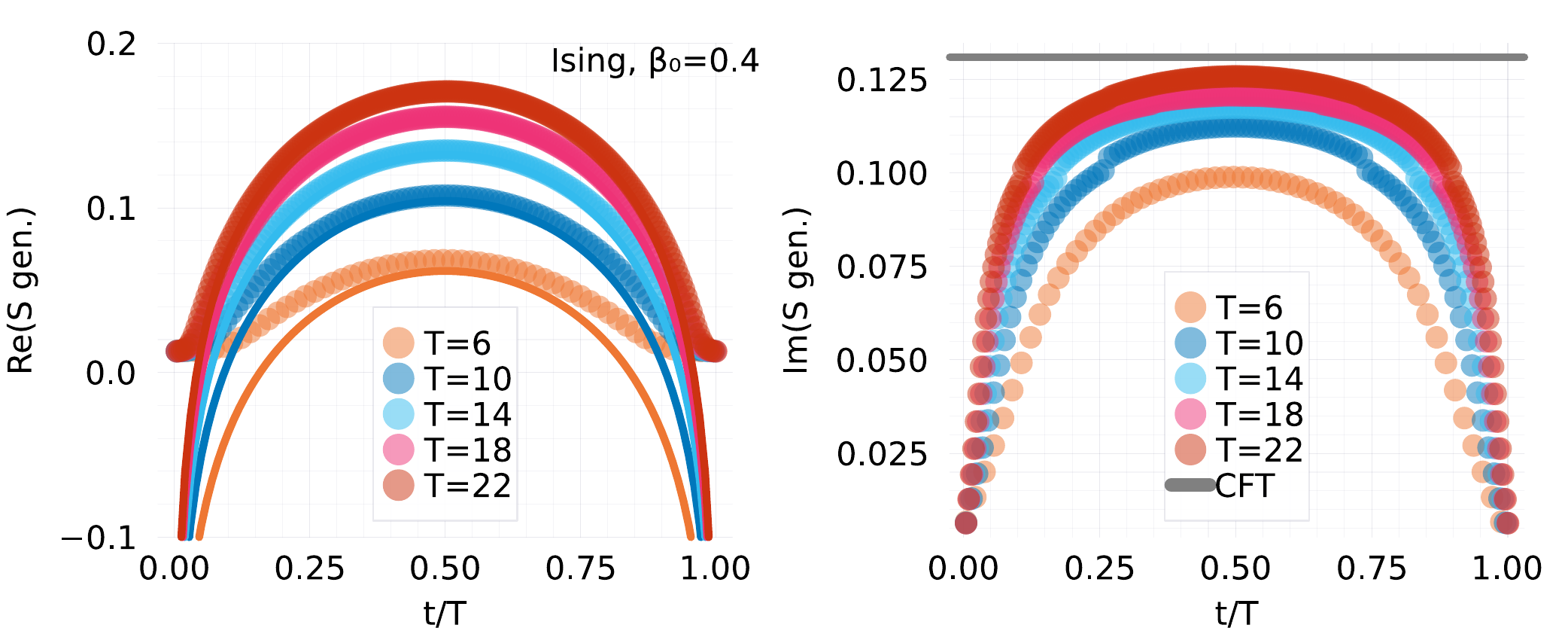}
 
  \includegraphics[width=.6\columnwidth]{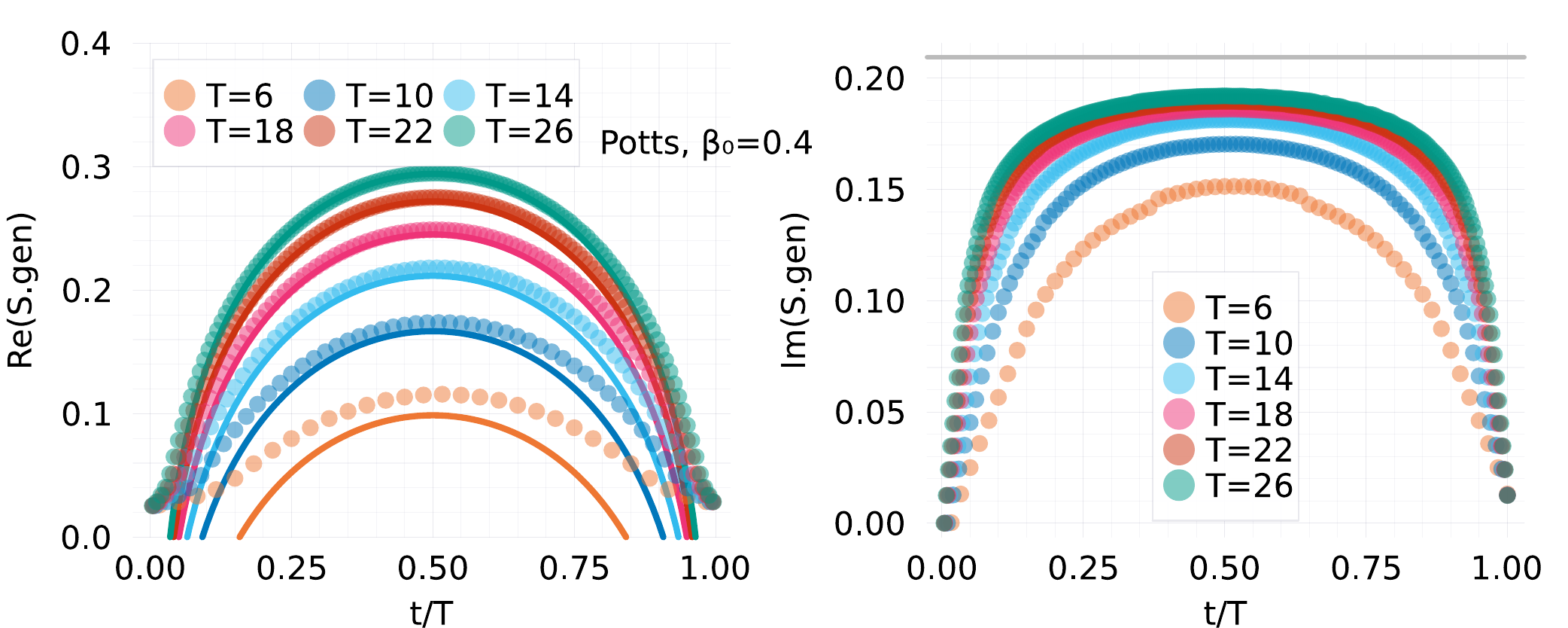}
\end{center}
 \caption{Numerical results for the real (left) and imaginary (right) parts of the generalized entropies for the critical Ising (top row) and Potts (bottom row) with
  fixed BC.
 Thick solid lines denote the CFT predictions.
  \label{fig:ents_slow}}
\end{figure}

The rate at which our results approach the CFT predictions of  the generalized entropies is affected by the choice of initial state for the \echo, which can be associated with different boundary CFT states and as a result with a different operator content of the theory. As an example, we show in Fig.~\ref{fig:ents_slow} the curves obtained for Ising and Potts using fixed BC. There we can appreciate that the convergence to the CFT results is significantly slower than the corresponding one for free BC, particularly for approach of  the imaginary part to the expected constant value.

%
 Such large deviations are quite usual in the context of studying the scaling of the entanglement entropy in different scenarios   \cite{cardy2010,alba2010,alba2011a,calabrese2013,coser2014a}.
  Here we show that even  for generalized entropies such corrections are induced by  finite size effects and vanish as inverse  powers of the IR and UV cutoff.
  We expect indeed that  both a finite  $\beta_0$ and $T$ to introduce  perturbations to our system.
  In the following, we will focus on the distance between the numerical results from our TN simulations for the imaginary part (taken at midchain) and the expected CFT result, ie. $\Delta S = Im(S_{gen})_{CFT} - Im(S_{gen})_{TN}$ even if one could work with the full chain length, in order to study the approach of all the points to the CFT results.
  
  In order to investigate the effect of $T$, we can use the standard  analysis described in \cite{cardy2010}:
  In particular, we expect that the leading corrections should scale as some inverse power of $\Delta S \propto T^{-x}$, with $x$ the lowest neutral relevant field.
  This implies that for Ising $x =1=\Delta_\epsilon$, while for Potts we expect $x= 4/5= \Delta_\epsilon$. 

\begin{figure}
\begin{center}
Ising fixed BC, $\gamma = 0.5, x= 1$

  \includegraphics[width=.28\columnwidth]{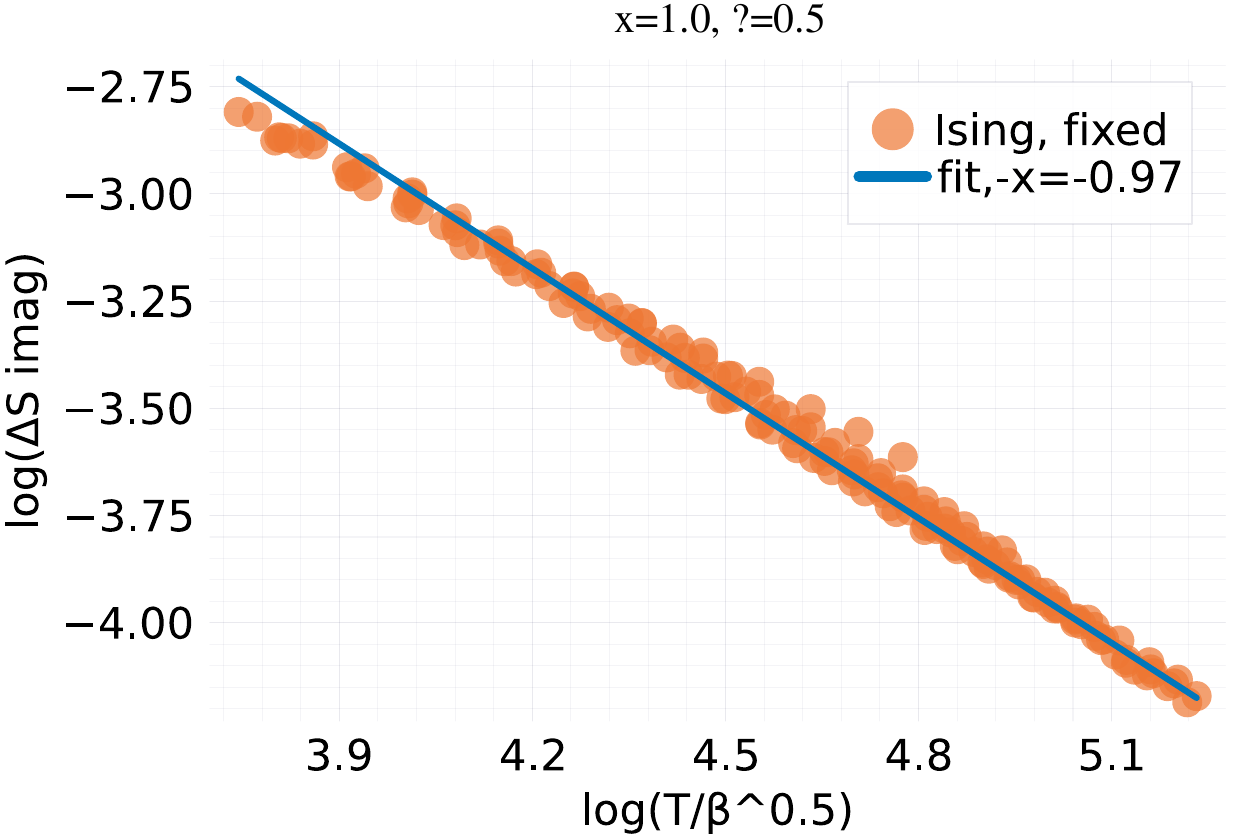}
  \includegraphics[width=.28\columnwidth]{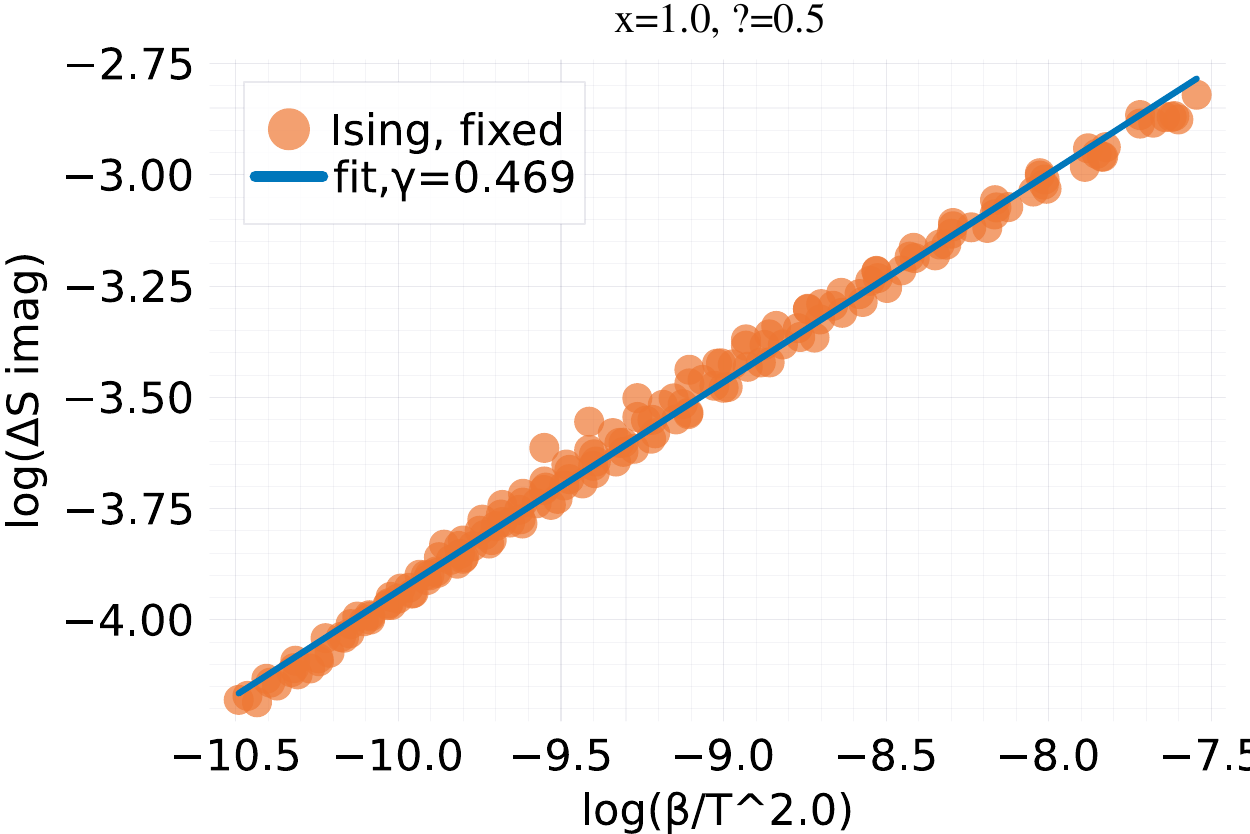}
  
  Ising free BC, $\gamma = 1.5, x= 1$
  
  \includegraphics[width=.28\columnwidth]{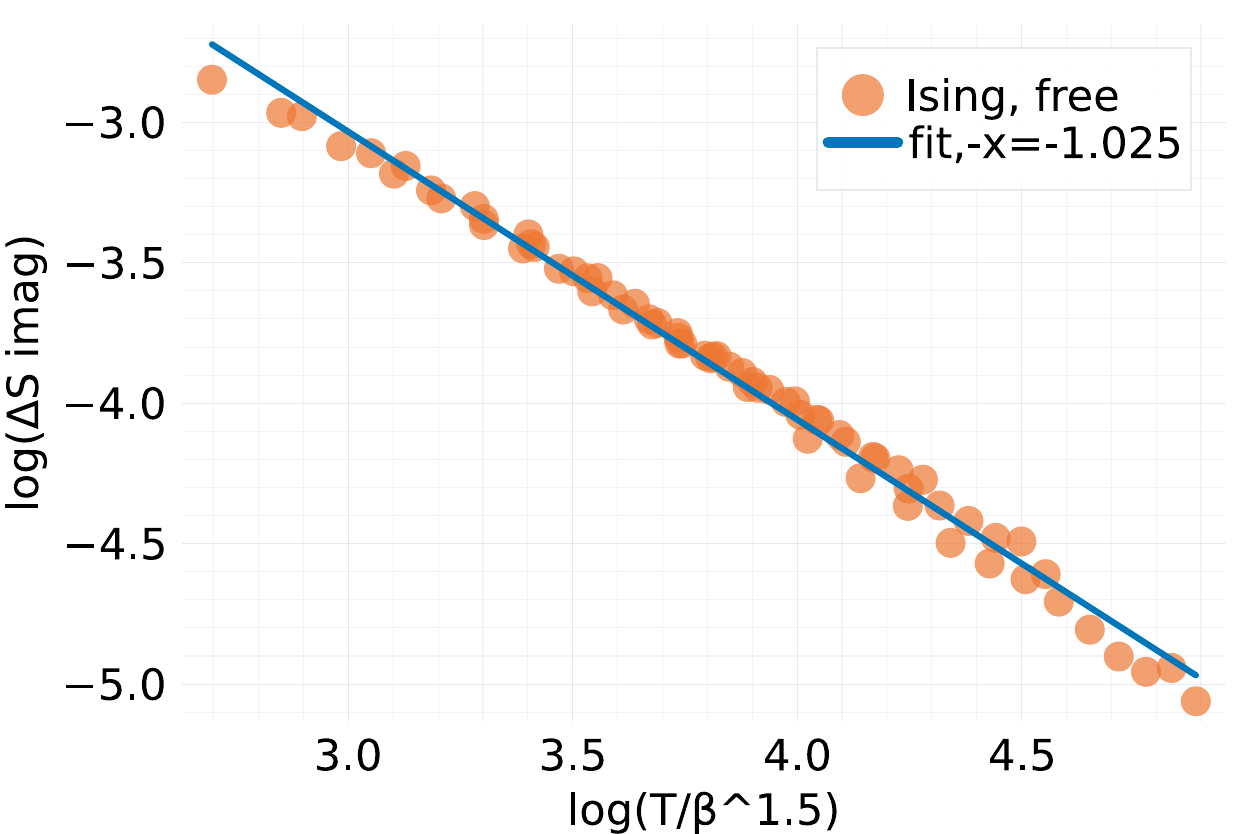}
  \includegraphics[width=.28\columnwidth]{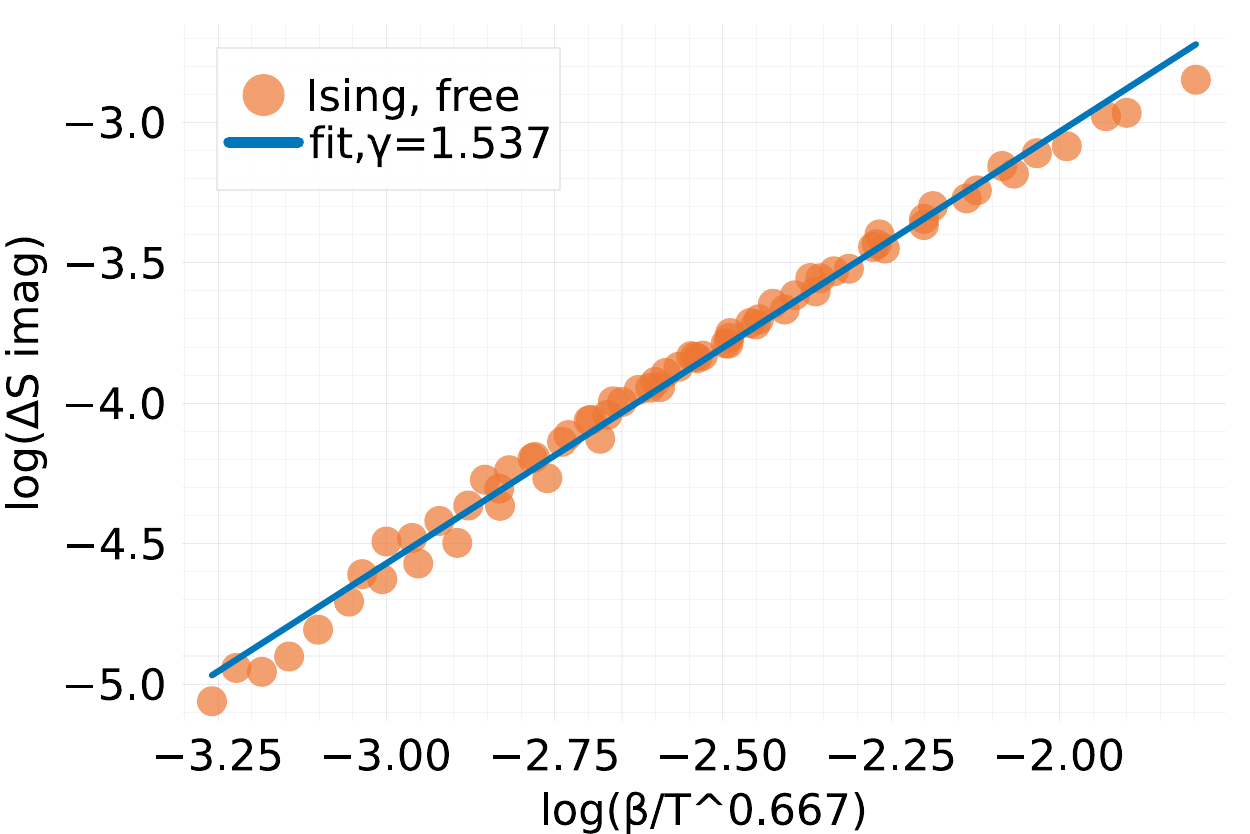}

  Potts fixed BC, $\gamma = 2/5, x = 4/5$
  
    \includegraphics[width=.28\columnwidth]{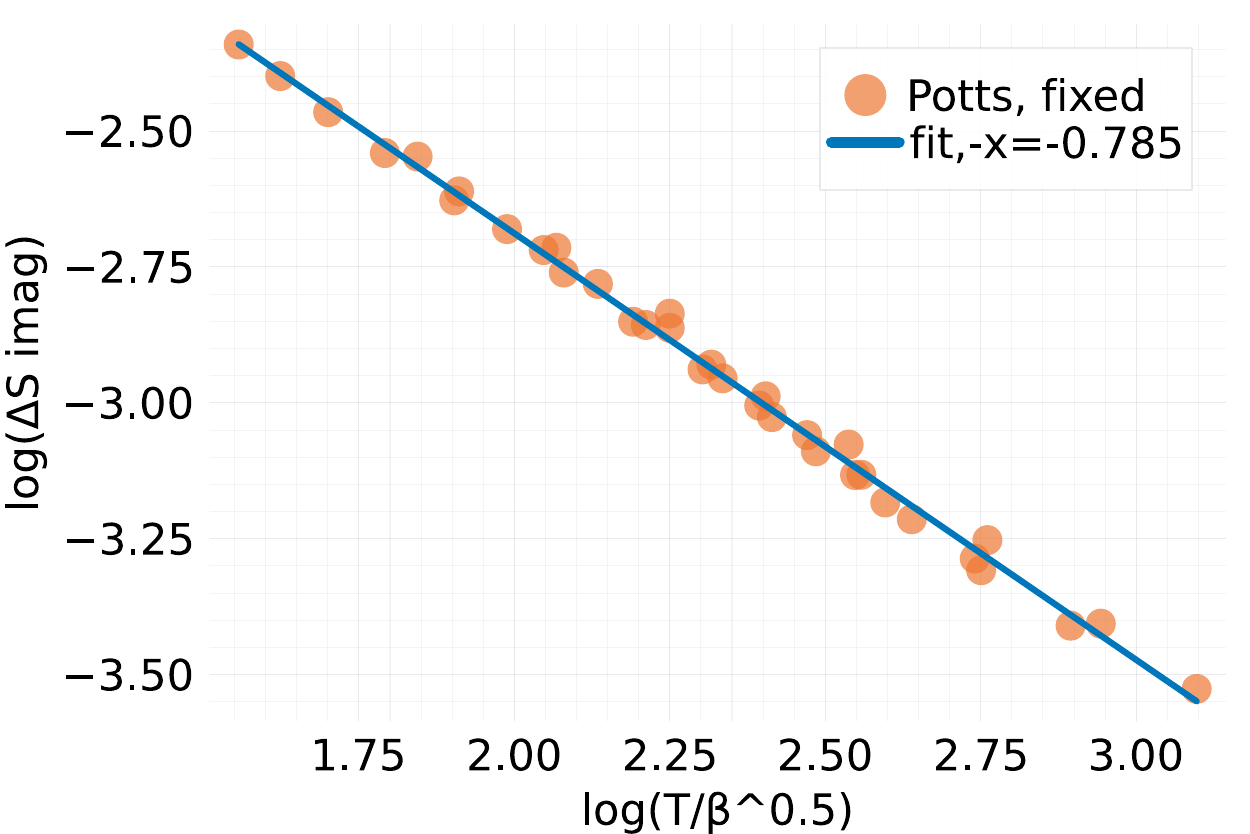}
  \includegraphics[width=.28\columnwidth]{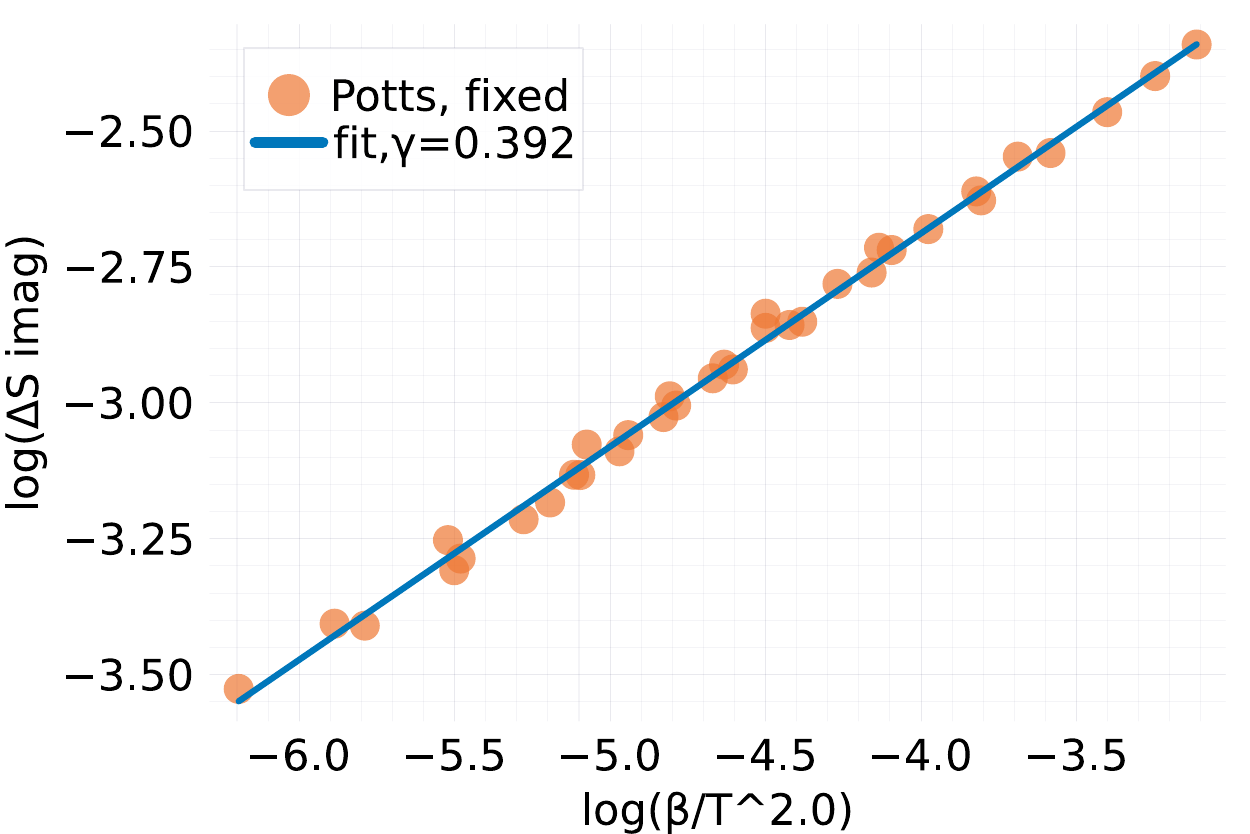}
  
   Potts free BC, $\gamma = 8/5, x = 4/3$
     
    \includegraphics[width=.28\columnwidth]{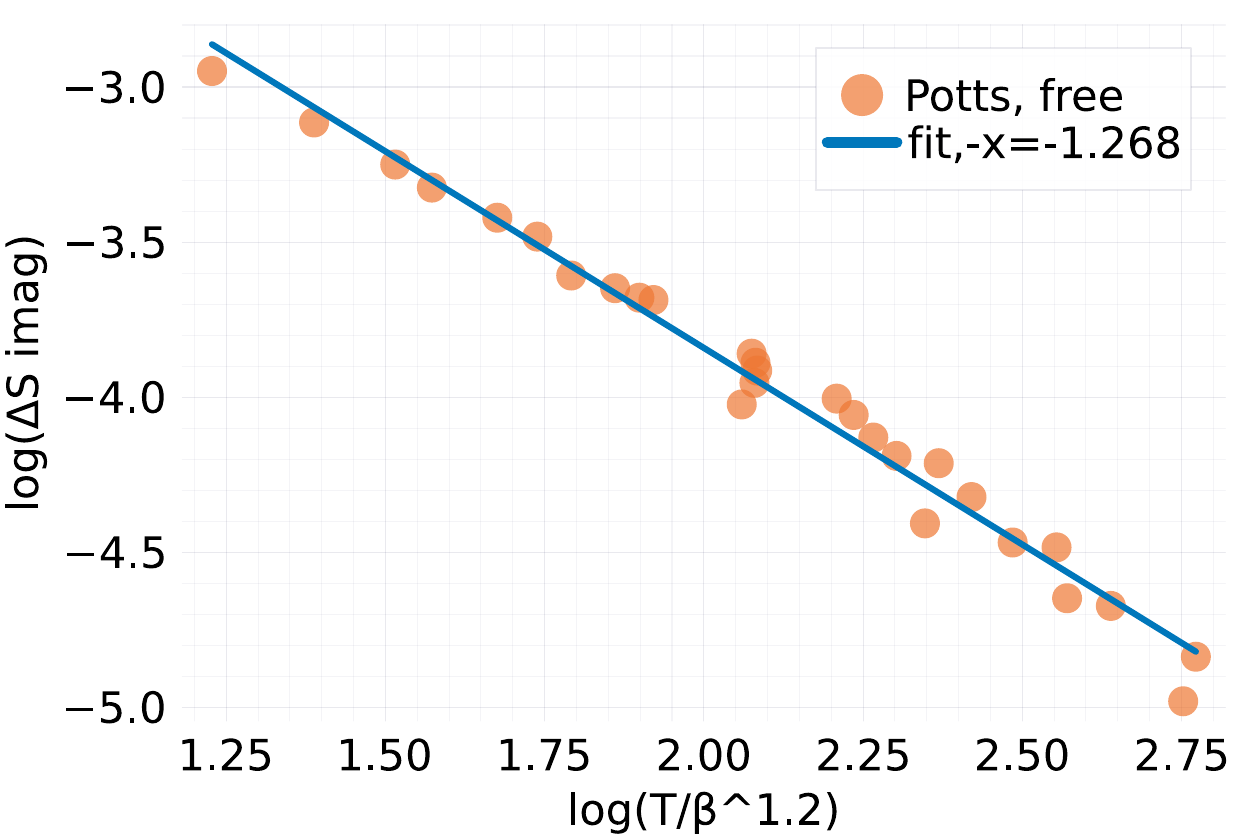}
  \includegraphics[width=.28\columnwidth]{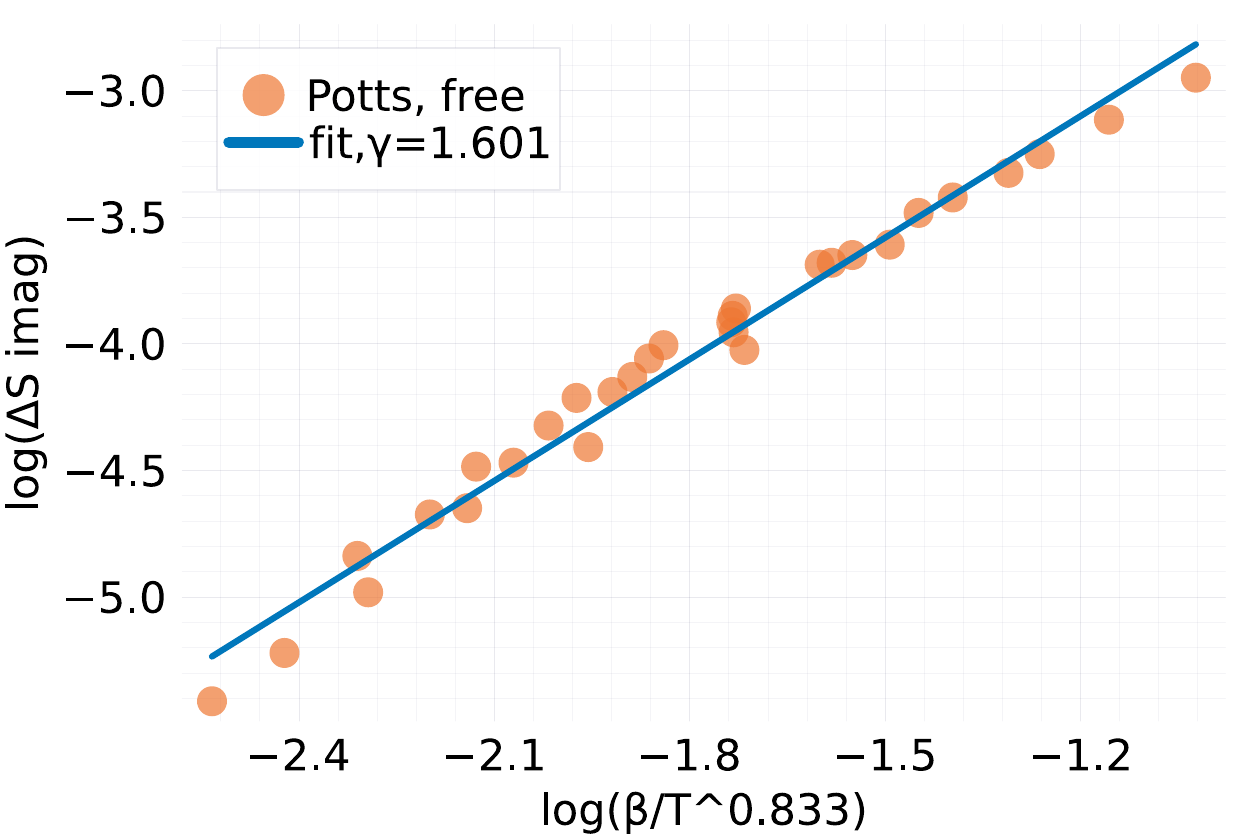}
  
  \end{center}
\caption{ Scaling behavior for $\Delta S$,  the difference of the imaginary part (at mid-chain) of the Ising (first rows) and Potts (bottom rows) entropies with the respect to the CFT predictions, for fixed and free boundary conditions.
$\Delta S$ for the various cases are plotted as function of $T/\beta_0^{\gamma/x}$ (left) and of
 $\beta_0/T^{x/\gamma}$ (right). The coefficients resulting from the best fits are shown in the plots. 
 \label{fig:scaling_ims_potts} }
\end{figure}

  {Our numerical results for Ising, as well as for Potts with fixed BC are indeed consistent with this kind of prediction, see  Fig.~\ref{fig:scaling_ims_potts}. }

Now we can add the effects of finite $\beta_0$: since $\beta_0$ plays the role of a UV cut-off, 
we now expect corrections $\Delta S \propto \beta_0 ^\gamma$ with $\gamma$ the scaling dimension of the most relevant operator coupling to $\Delta S$.
Putting everything together we thus expect that the scaling variable for the correction should be  $\beta_0^{\gamma/x}/{T}$.

{
Performing our fits, we find that for Ising with free BC an exponent $\gamma =1.5$ which is compatible with $\gamma=2-\Delta'_\sigma$, where $\Delta'_\sigma=0.5$ is the boundary magnetization. For fixed BC on the other hand our best fit suggests $\gamma=0.5$.
For Potts with fixed BC initial state we find instead $\gamma=2/5$, which once more is compatible with $\Delta_\epsilon$.  The exponents change as expected  for the Potts model with free BC: there the best fits give
an exponent $x \simeq 4/3$ (compatible with a neutral para-fermion condensate) and $\gamma\simeq 8/5$, which is compatible with $\gamma=2-2/5$,
see Fig.~\ref{fig:scaling_ims_potts}
}


\end{appendix}

\end{document}